\documentstyle[12pt,amsmath,amssymb,epic]{article}

\newcommand{\bbbone}{\mathchoice {\rm 1\mskip-4mu l} {\rm 1\mskip-4mu l}
{\rm 1\mskip-4.5mu l} {\rm 1\mskip-5mu l}}
\newcommand{\scalprod}[2]{\left\langle {#1}, {#2}\right\rangle}
\newcommand{\dom}{{\cal D}}
\newcommand{\RE}{{\rm Re}}
\newcommand{\IM}{{\rm Im}}
\newcommand{\fer}[1]{(\ref{#1})}
\newcommand{\ran}{{\rm Ran\,}}
\newcommand{\repsilonbar}{\,\overline{\!R}_\epsilon}
\newcommand{\repsilon}{R_\epsilon}

\newcommand{\h}{{\cal H}}
\newcommand{\cx}{{\mathbb C}}
\newcommand{\r}{{\mathbb R}}

\newcommand{\supp}{{\rm supp}}

\renewcommand{\d}{{\rm d}}
\newcommand{\Pibar}{\overline{\Pi}}
\newcommand{\Pbar}{\overline{P}}

\newcommand{\dirint}{\int^\oplus}

\newcommand{\av}[1]{\left\langle{#1}\right\rangle}
\newcommand{\hh}{{\frak H}}
\newcommand{\What}{\what{W}}

\newcommand{\pbar}{\,\overline{p}\,}
\newcommand{\mm}{{\frak M}}
\newcommand{\pp}{{\cal P}}
\newcommand{\cc}{{\cal C}}

\newcommand{\alphaepsilon}{\alpha_{t,\lambda}^{(\epsilon)}}
\newcommand{\sigmaepsilon}{\sigma_{t,\lambda}^{(\epsilon)}}
\newcommand{\oref}{\omega_{\rho_0,\beta}}
\newcommand{\Oref}{\Omega_{\rho_0}}

\renewcommand{\O}[1]{O\left({#1}\right)}

\newcommand{\what}[1]{\widehat{#1}}
\newcommand{\Mbar}{\,\overline{\!M}}
\renewcommand{\tilde}[1]{\widetilde{#1}}

\newcounter{resultcounter} 
\stepcounter{resultcounter}

\newtheorem{theorem}{Theorem}[section]

\newtheorem{proposition}{Proposition}[section]

\newcommand{\rep}{{\rm Rep\,}}
\newcommand{\ww}{{\frak W}}

\begin{document}

\setcounter{page}{0}

\title{Thermal Ionization}

\author{
J\"urg Fr\"{o}hlich\footnote{juerg@itp.phys.ethz.ch}\ \ \  and \  
Marco Merkli\footnote{merkli@itp.phys.ethz.ch}\ \footnote{Supported by an NSERC
  Postoctoral Fellowship and by ETH-Z\"urich}\\      
Theoretical Physics\\
 ETH- H\"{o}nggerberg \\ 
CH-8093 Z\"{u}rich, Switzerland\\
}
\date{\today\\
\ \\
\small \em Dedicated, in admiration and friendship, to\\
Elliott Lieb and Edward Nelson,\\
 on the occasion of their $70^{\rm th}$ birthdays. }
\maketitle

\begin{abstract}
In the context of an idealized model describing an atom coupled to black-body
radiation at a sufficiently high positive temperature, we show that the
atom will end up being ionized in the limit of large times. Mathematically,
this is translated into the statement that the coupled system does not have
any time-translation invariant state of positive (asymptotic) temperature, and
that the expectation value of an arbitrary finite-dimensional projection in an
arbitrary initial state of positive (asymptotic) temperature tends to zero, as
time tends to infinity. \\
\indent
These results are formulated within the general framework of $W^*$-dynamical
systems, and the proofs are based on Mourre's theory of positive commutators
and a new virial theorem. Results on the so-called standard form of a von
Neumann algebra play an important role in our analysis. 
\end{abstract}
\thispagestyle{empty}
\setcounter{page}{1}
\setcounter{section}{1}


\setcounter{section}{0}

\section{Introduction}

In this paper, we study an idealized model describing an atom or molecule
consisting of static nuclei and electrons coupled to black-body radiation. Our
aim is to show that when the quantized radiation field is in a thermal state
corresponding to a sufficiently high positive temperature, and under
suitable conditions on the interaction Hamiltonian, including infrared and
ultraviolet cutoffs and a small value of the coupling constant, the atom or
molecule will always be ionized in the limit of very large times. This process
is called {\it thermal ionization}. \\
\indent
Thus, a very dilute gas of atoms or molecules in intergalactic space and
subject to the $3K$ thermal background radiation of the universe will
eventually be transformed into a very dilute plasma of nuclei and electrons.\\
\indent
If the temperature of the black-body radiation is small, as compared to a
typical atomic ionization energy, then an atom initially prepared in an
excited bound state will start to emit light and relax towards its
groundstate. After a time much longer than its relaxation time, it will be
stripped of its electrons in very unlikely events where an atomic electron is
hit by a high-energy photon from the thermal background radiation. The life
time of the groundstate of an isolated atom interacting with black body
radiation at inverse temperature $\beta$, before it is ionized, is expected to
be exponentially large in $\beta$. A precise
description of the temporal evolution of such an atom is difficult to come by;
but the claim that, it will eventually be
ionized, is highly plausile. To most physicists, this result must look
obvious. Unfortunately a complete 
proof of it is likely to be very involved. The main purpose of this paper is to
present some partial results, thermal ionization at sufficiently high
temperatures for simplified models, supporting this picture.\\
\indent
If the temperature of electromagnetic radiation is strictly zero then an atom
initially prepared in a bound state of maximal energy well below its
ionization threshold can be shown to always relax to a groundstate by emitting
photons; (for a proof of this statement in some slightly idealized models see
[FGS]). This result and our complementary result on thermal ionization provide
some qualitative understanding of two fundamental irreversible processes in
atomic physics: relaxation to a ground state, and ionization by thermal
radiation.\\
\indent
Next, we describe the physical system analyzed in this paper somewhat more
precisely; (for further details see Section \ref{defmod}). It is composed of a
subsystem with finitely many degrees of freedom, the ``atom'' (or
``molecule''), and a subsystem with infinitely many degrees of freedom, the
``radiation field''. The space of pure state vectors of the atom is a
separable Hilbert space, $\h_p$; (where the subscript $_p$ stands for
``particle''). Mixed states of the atom are described by density matrices,
$\rho$, where $\rho$ is a non-negative, trace-class operator on $\h_p$ of unit
trace. The expectation value of a bounded operator $A$ on $\h_p$ in the state
$\rho$ is given by 
\begin{equation}
\omega_\rho^p(A):=\mbox{tr\,} \rho A.
\label{j1}
\end{equation}
\indent
Before the ``atom'' or particle system is coupled to the radiation field the
time evolution of a bounded operator $A$ on $\h_p$ in the Heisenberg picture
is given by 
\begin{equation}
\alpha_t^p(A):= e^{itH_p} Ae^{-itH_p},
\label{j3}
\end{equation}
where $H_p$ is the particle Hamiltonian, which is a selfadjoint operator on
$\h_p$ whose spectrum is bounded from below by a constant $E>-\infty$. \\
\indent
To be specific, we may think of $\h_p$ as being the Hilbert space 
\begin{equation}
\h_p={\mathbb C}^n\oplus L^2(\r^3, d^3x),
\label{j4}
\end{equation}
and the Hamiltonian $H_p$ as the operator
\begin{equation}
H_p=\mbox{diag\,}(E_0=E, E_1,\ldots,
  E_{n-1})\upharpoonright_{\cx^n} \oplus
  (-\Delta)\upharpoonright_{L^2(\r^3,d^3x)},
\label{j5}
\end{equation}
describing a one-electron atom (with a static nucleus) with $n$ boundstates of
energies $E_0, E_1,\ldots,E_{n-1}<0$ and scattering states of arbitrary
energies 
$k^2\in[0,\infty)$ spanning the subspace $L^2(\r^3,d^3x)$ of $\h_p$. Thus, the
point spectrum of $H_p$ is given by the eigenvalues $\{E_0, E_1,\ldots,
E_{n-1}\}$ and the continuous spectrum of $H_p$ covers $[0,\infty)$, has
constant (infinite) multiplicity and is absolutely continuous. Just in order
to keep things simple, we shall usually assume that $n=1$. \\
\indent
The bounded operators on a Hilbert space $\h$ form a von Neumann algebra
denoted by ${\cal B}(\h)$. A convenient algebra of operators encoding the
kinematics of the ``atom'' or particle sytem is the algebra ${\frak A}_p:={\cal
  B}(\h_p)$.\\ 
\indent
The ``radiation field'' is described by a free, massless, scalar Bose field
$\varphi$ on physical space $\r^3$, a ``phonon field''. For purposes of
physics, it would be preferable to replace $\varphi$ by the free
electromagnetic field. In our entire analysis, this replacement can be made
without any difficulties - at the price of slightly more complicated
notation. A convenient algebra of operators to encode the kinematics of the
radiation field is a $C^*$-algebra ${\frak A}_f$ which can be viewed as a
time-averaged version of the algebra of Weyl operators generated by $\varphi$
and its conjugate momentum field $\pi$. The time
evolution of operators in ${\frak A}_f$, in the Heisenberg picture, {\it
  before} the field is coupled to the particle system, is given by the {\it
  free-field time evolution} $\alpha_t^f$, which is a one-parameter 
group of $*$automorphisms on ${\frak A}_f$. \\
\indent
A one-parameter group $\{\alpha_t | t\in\r\}$ defined on a $C^*$-algebra
$\frak A$ is a $*$auto\-morphism group of $\frak A$ iff 
\begin{eqnarray}
\alpha_t(A)&\in& {\frak A}, \ \ 
(\alpha_t(A))^*=\alpha_t(A^*),\mbox{\ \ for all
  $A\in\frak A$,}\nonumber\\
\alpha_t(A)\alpha_t(B)&=&\alpha_t(AB),\mbox{\ \ for all
  $A, B\in\frak A$},\nonumber\\
\alpha_{t=0}(A)&=&A,\ \ \alpha_t(\alpha_s(A))=\alpha_{t+s}(A),\mbox{\ \ for all
  $A\in\frak A$, $t,s\in\r$.}
\label{j6}
\end{eqnarray}
Since we work on a time-averaged Weyl algebra, the free field time evolution is
{\it norm continuous}, i.e., $t\mapsto \alpha^f_t(A)$ is a continuous map from
$\r$ to ${\frak A}_f$.  
General states of the radiation field can be described as states on the
algebra ${\frak A}_f$, i.e., as positive, linear functionals, $\omega$, on
${\frak A}_f$ normalized such that $\omega(\bbbone)=1$.\\
\indent
A convenient algebra of operators to encode the kinematics of the system
composed of the ``atom'' and the ``radiation field'' is the $C^*$-algebra,
${\frak A}$, given by 
\begin{equation}
{\frak A}={\frak A}_p\otimes {\frak A}_f.
\label{j7}
\end{equation}
The time evolution of operators in $\frak A$, {\it before} the two subsystems
are coupled to each other, is given by 
\begin{equation}
\alpha_{t, 0}:=\alpha_t^p\otimes \alpha_t^f.
\label{j8}
\end{equation}
\indent
A regularized interaction coupling the two subsystems can be introduced by
choosing a bounded, selfadjoint operator $V^{(\epsilon)}\in{\frak A}$, where the
superscript $^{(\epsilon)}$ indicates that a regularization has been imposed on an
interaction term, $V$, in such a way that $\|V^{(\epsilon)}\|=\O{1/\epsilon}$. We
define the {\it regularized, interacting time evolution of the coupled system}
as a $*$automorphism group $\{\alpha_{t,\lambda}^{(\epsilon)} | t\in\r\}$ on
the algebra $\frak A$ given by the norm-convergent {\it Schwinger-Dyson
  series}
\begin{eqnarray}
\lefteqn{
\alpha_{t,\lambda}^{(\epsilon)}(A)=\alpha_{t,0}(A) +\sum_{n=1}^\infty
(i\lambda)^n \int_0^t dt_1\cdots}\nonumber\\
&&\cdots\int_0^{t_{n-1}} dt_n\ 
[\alpha_{t_n,0}(V^{(\epsilon)}),[\alpha_{t_{n-1},0}(V^{(\epsilon)}),\ldots,
[\alpha_{t_1,0}(V^{(\epsilon)}),\alpha_{t,0}(A)]\cdots]],\ \ \ \ \ 
\label{j9}
\end{eqnarray} 
for an arbitrary operator $A\in{\frak A}$. In equation \fer{j9}, $\lambda$ is
a coupling constant, and the interaction term $V$ is chosen in accordance with
conventional models describing electrons coupled to the quantized radiation
field. \\
\indent
We are interested in analyzing the time evolution of the coupled system in
some states $\omega$ of physical interest, i.e., in understanding the
time-dependence of expectation values
\begin{equation}
\omega(\alpha_{t,\lambda}^{(\epsilon)}(A)),\ \ A\in{\frak A},
\label{j10}
\end{equation}
in the limit where the regularization is removed, i.e., $\epsilon\rightarrow
0$, and for large times $t$. The {\it states} $\omega$ {\it of interest} are
states ``close to'' (technically speaking, {\it normal} with respect to) a {\it
  reference state} of the form 
\begin{equation}
\omega_{\rho,\beta}:=\omega_\rho^p\otimes\omega_\beta^f,
\label{j11}
\end{equation}
where $\omega_\rho^p$ is given by a density matrix $\rho$ on $\h_p$, see
equation \fer{j1}, and $\omega_\beta^f$ is the {\it thermal equilibrium state}
of the radiation field at temperature $T=(k_B\beta)^{-1}$, where $k_B$ is
Boltzmann's constant. Technically, $\omega_\beta^f$ is defined as the {\it
  unique} $(\alpha_t^f,\beta)$-KMS {\it state} on the algebra ${\frak A}_f$;
  it is invariant under (or ``stationary'' for) the free-field time evolution
  $\alpha_t^f$. If the density matrix $\rho$ describes an arbitrary
  statistical mixture of bound states of $H_p$, but $\rho$ vanishes on the
  subspace $L^2(\r^3,d^3x)$ of $\h_p$, then $\omega_{\rho,\beta}$ is {\it
    stationary} for the free time evolution $\alpha_{t,0}$ defined in equation
  \fer{j8}. However, it is {\it not} an equilibrium (KMS) state for
  $\alpha_{t,0}$. In fact, because $H_p$ has continuous spectrum, there are
  {\it 
    no} equilibrium (KMS) states on $\frak A$ for the time evolution
  $\alpha_{t,0}$.\\
\indent
Given the algebra $\frak A$ and a reference state $\omega_{\rho,\beta}$ on
$\frak A$, as in equation \fer{j11}, the {\it GNS construction} associates
with the 
pair $({\frak A},\omega_{\rho,\beta})$ a Hilbert space $\h$, a $*$representation
$\pi_\beta$ of $\frak A$ on $\h$, and a vector
$\Omega_\rho\in\h$, cyclic for the algebra $\pi_\beta({\frak
  A})$, such that 
\begin{equation}
\omega_{\rho,\beta}(A)=\scalprod{\Omega_{\rho}}{\pi_\beta(A)\Omega_{\rho}},
\label{j12}
\end{equation}
for all $A\in\frak A$. The closure of the algebra $\pi_\beta({\frak A})$ in
the weak operator topology is a von Neumann algebra of bounded operators on
$\h$ which we denote by $\mm_\beta$. This algebra depends on $\beta$,
but is independent of the choice of the density matrix $\rho$. The states
$\omega$ on $\frak A$ of interest to us are given by vectors $\psi\in\h$
in such a way that 
\begin{equation}
\omega(A)=\scalprod{\psi}{\pi_\beta(A)\psi},\ \ \ A\in\frak A.
\label{j13}
\end{equation}
\indent
We shall see that there exists a selfadjoint operator $L_\lambda^{(\epsilon)}$
on $\h$ generating the time evolution of the coupled system, in the
sense that 
\begin{equation}
\pi_\beta(\alpha_{t,\lambda}^{(\epsilon)}(A)) =e^{it L_\lambda^{(\epsilon)}}
\pi_\beta(A) e^{-itL_\lambda^{(\epsilon)}},
\label{j14}
\end{equation}
for $A\in\frak A$; $L_\lambda^{(\epsilon)}$ is called the (regularized) {\it
  Liouvillian}. Clearly,
\begin{equation}
\sigma_{t,\lambda}^{(\epsilon)}(K):= e^{it L_\lambda^{(\epsilon)}}
K e^{-itL_\lambda^{(\epsilon)}}, \ \ K\in\mm_\beta,
\label{j15}
\end{equation}
defines a $*$automorphism group of time translations on $\mm_\beta$. For an
interesting class of models, we shall show that 
\begin{equation}
\mbox{s-}\lim_{\epsilon\rightarrow 0} e^{itL_\lambda^{(\epsilon)}} =:
e^{itL_\lambda}
\label{j16}
\end{equation}
exists, for all $t$, and defines a unitary one-parameter group on
$\h$. It then follows from \fer{j15}, \fer{j16} that 
\begin{equation}
\sigma_{t,\lambda}(K):=e^{itL_\lambda} K e^{-itL_\lambda}
\label{j17}
\end{equation}
defines a one-parameter group of $*$automorphisms on the von Neumann algebra
$\mm_\beta$. The pair $(\mm_\beta,\sigma_{t,\lambda})$ defines a so-called
{\it $W^*$-dynamical system}. If the coupling constant $\lambda$ vanishes then
a state $\omega_{\rho,\beta}=\omega_\rho^p\otimes\omega_\beta^f$, where the density
matrix $\rho$ vanishes on the subspace $L^2(\r^3,d^3x)\subset\h_p$
corresponding to the continuous spectrum of $\h_p$ and commutes with $H_p$, is
an invariant state for $\sigma_{t,0}$, in the sense that 
\begin{equation}
\omega_{\rho,\beta}(\sigma_{t,0}(K)):=
\scalprod{\Omega_{\rho}}{\sigma_{t,0}(K)\Omega_{\rho}}=\omega_{\rho,\beta}(K),
\label{j18}
\end{equation}
for all $K\in\mm_\beta$. \\
\indent
The {\it main result} proven in this paper can be described as follows: For an
interesting class of interactions, $V$, for an arbitrary inverse temperature
$0<\beta<\infty$, and for all real coupling constants
$\lambda$ with $0<|\lambda|<\lambda_0(\beta)$, where $\lambda_0(\beta)$
 depends on the choice of $V$, and on $\beta$ as $\lambda_0(\beta)\sim
 e^{\beta E_0}$, where $E_0<0$ is the ground state energy of the particle
 system, there {\it do not exist 
  any} states $\omega$ on $\mm_\beta$ close, in the sense of equation
\fer{j13}, to a reference state $\omega_{\rho,\beta}$, as in equation
\fer{j11}, 
which are {\it invariant} under the time evolution $\sigma_{t,\lambda}$ on
$\mm_\beta$, (in the sense that $\omega(\sigma_{t,\lambda}(K))=\omega(K)$, for
$K\in\mm_\beta$). \\
\indent
In other words, we show that, under the hypotheses described above, there
are no time-translation invariant states of the coupled system of asymptotic
temperature $T=(k_B\beta)^{-1}>0$. It will turn out that this result is a
consequence of the following one: For a certain canonical definition of the
Liouvillian $L_\lambda$ of the coupled system, and under the hypotheses
sketched above, $L_\lambda$ does not have any eigenvectors $\psi\in\h$, in particular, $\mbox{ker}L_\lambda=\{0\}$. This result will be proven with the help of
{\it Mourre's theory of positive commutators} applied to $L_\lambda$ and a new
virial theorem.\\
\indent
As a corollary of our results it follows that, for an arbitrary vector
$\psi\in\h$ and an arbitrary compact operator $K$ on $\h$, 
\begin{equation}
\scalprod{\psi}{e^{itL_\lambda}K e^{-itL_\lambda}\psi}\rightarrow 0,
\label{j19}
\end{equation}
as time $t\rightarrow \infty$, (at least in the sense of ergodic means). This
means that the survival probability of an arbitrary bound state of the atom
coupled to the quantized radiation field in a thermal equilibrium state at
positive temperature tends to zero, as time $t\rightarrow
\infty$. Heuristically, this can be understood by using Fermi's Golden
Rule. \\
\indent
One may wonder how the quantum-mechanical motion of an electron looks like,
after it has been knocked off the atom by a high-energy boson, i.e., after
thermal ionization. We cannot give an answer to this question, in this paper,
because we are not able to analyze appropriately realistic models, yet. But it
is natural to expect that this motion will be {\it diffusive}, furnishing an
example of ``quantum Brownian motion''. Progress on this question would be
highly desirable.\\

{\it Organization of the paper.\ } In Section \ref{ti}, we define the model,
and state our main result on thermal ionization, 
Theorem \ref{2thm}, which follows from spectral properties of the
Liouvillian proven in our key technical theorem, Theorem \ref{mainthm}. In
Section \ref{vtsect}, we 
state two general virial theorems, Theorems \ref{virialthm} and
\ref{virialthm'}, we present a result on regularity of
eigenfunctions of Liouvillians, Theorem \ref{roethm}, and explain some basic
ideas 
of the positive commutator method. The proof of Theorem \ref{mainthm}
(spectrum of Liouvillian) is given in Section \ref{mainthmproofsect}. It
consists of two main parts:  verification that the virial theorems are
applicable in the particular situation encountered in the analysis of our
models (Subsection \ref{concsetting}), and
proof of a lower bound on a commutator of the Liouvillian with a suitable
conjugate operator (Subsections \ref{masection},
\ref{feshsection}). In Section \ref{sfa}, we establish some technical results
on the 
invariance of operator domains and on certain commutator expansions that are needed in the proofs
of the virial theorems and of the theorem on regularity of
eigenfunctions. Proofs of the latter results are presented in Section
\ref{vtproof}. In Section \ref{flowsection}, we 
describe some 
results on unitary groups generated by vector fields which are needed in the
definition of our ``conjugate operator'' $A^a_p$ in the positive commutator
method. The last section, Section \ref{propproofsection}, contains proofs of
several propositions used in earlier sections of the paper.

\section{Definition of models and main results on thermal ionization}
\label{ti}

In Section \ref{defmod}, we introduce our model and use it to define a
$W^*$-dynamical system $(\mm_\beta,\sigma_{t,\lambda})$. Our main results on
thermal ionization are described in Section \ref{tisect}.

\subsection{Definition of the model}
\label{defmod}
Starting with the algebra $\frak A$ and a (regularized)
dynamics $\alphaepsilon$ on it, we introduce a reference state $\oref$, and
consider the induced (regularized)
dynamics $\sigmaepsilon$ on $\pi_\beta(\frak A)$, where $(\h,\pi_\beta,\Oref)$
denotes the GNS representation corresponding to $({\frak A},\oref)$.  We show that, 
as 
$\epsilon\rightarrow 0$, $\sigmaepsilon$ tends 
to a $*$automorphism group, $\sigma_{t,\lambda}$, of the von Neumann algebra
$\mm_\beta$, defined as the weak closure of $\pi_\beta(\frak A)$ in ${\cal
  B}(\h)$. We determine the generator, $L_\lambda$, of the unitary group, $e^{itL_\lambda}$, on
$\h$ implementing $\sigma_{t,\lambda}$; $L_\lambda$ is called a 
{\it Liouvillian}. The relation between
eigenvalues of $L_\lambda$ and invariant normal states on $\mm_\beta$ will be
explained later in this section, (see Theorem \ref{nostatstatesthm}). We will
sometimes write simply $L$ instead of $L_\lambda$, for $\lambda\neq 0$.

\subsubsection{The algebra ${\frak A}_f$}
We introduce a $C^*$-algebra suitable for the description of the dynamics of
the free field, and, as we explain below, for the description of the
interacting dynamics. \\
\indent
Let ${\frak W}={\frak W}(L^2_0)$ be the Weyl CCR algebra over
\begin{equation*}
L^2_0:=L^2(\r^3,d^3k)\cap L^2(\r^3,|k|^{-1}d^3k),
\end{equation*}
i.e., the $C^*$-algebra generated by the Weyl operators, $W(f)$, for $f\in
L^2_0$, satisfying
\begin{equation*}
W(-f)=W(f)^*,\ \ \ W(f)W(g)=e^{-i\IM \scalprod{f}{g}/2}W(f+g).
\end{equation*}
Here, $\scalprod{\cdot}{\cdot}$ denotes the inner product of $L^2_0$. The
latter relation implies the CCR
\begin{equation}
W(f)W(g)=e^{-i\IM\scalprod{f}{g}}W(g)W(f).
\label{CCR}
\end{equation}
\indent
The expectation functional for the KMS state 
of an infinitely extended free Bose field in thermal equilibrium at
inverse temperature $\beta$ is given by
\begin{equation*}
g\mapsto \omega^f_\beta\left(W(g)\right)=\exp\left\{
  -\frac{1}{4}\int_{\r^3}\left( 1+\frac{2}{e^{\beta |k|}-1}\right) |g(k)|^2
  d^3k\right\},
\end{equation*}
which motivates the choice of the space $L^2_0$ (as opposed to $g\in
L^2(\r^3)$). \\
\indent 
The free field dynamics on $\ww$ is given by the $*$automorphism group
\begin{equation}
\alpha_t^\ww(W(f))=W(e^{i|k|t}f).
\label{weyldyn}
\end{equation}
It is well known that for $f\neq 0$, $t \mapsto \alpha^\ww_t(W(f))$ is not a
continuous map from $\r$ to $\ww$, but $t\mapsto \omega(\alpha^\ww_t(W(f)))$
is continuous for a large (weak* dense) class of states $\omega$ on
$\ww$. An interacting dynamics is commonly defined using a Dyson series
expansion, hence we should be able to give a sense to time integrals over
 $\alpha^\ww_t(a)$, for $a\in\ww$. Because of the lack of
 norm-continuity of the free dynamics, such an integral cannot be interpreted
 in norm sense, but only in a weak hence representation
 dependent way. In order to give a representation independent definition of the
 (coupled) dynamics, we modify the algebra in such a way that the free
 dynamics becomes norm continuous. The idea is to introduce a time-averaged 
 Weyl algebra, generated by elements given by
\begin{equation}
a(h)=\int_\r ds\ h(s)\alpha_s^\ww(a),
\label{ah}
\end{equation}
for functions $h$ in a certain class, and $a\in\ww$ (if $h$ is sharply
localized at zero, the integral approximates $a\in\ww$). The free dynamics is
then given by
\begin{equation*}
\alpha_t^f(a(h))=\int_\r ds\ h(s) \alpha_s^\ww(\alpha_t^\ww(a)) =\int_\r ds\
h(s-t)\alpha_s^\ww(a).
\end{equation*}
We now construct a
$C^*$-algebra  
whose elements, when represented on a Hilbert space, are given by \fer{ah},
where the integral is understood in a weak sense.\\

Let $\frak P$ be the free algebra generated by elements
\begin{equation*}
\left\{a(h)\ |\ a\in{\frak W}, \widehat{h}\in C_0^\infty(\r)\right\},
\end{equation*}
where $\, \widehat{\, }$ denotes the Fourier transform. Taking the functions
$h$ to be analytic (i.e., having a Fourier transform in $C^\infty_0$) allows us
to construct KMS states w.r.t. the free dynamics,  as we explain below.  
We equip the algebra $\frak P$ with the star operation defined by
$(a(h))^*=(a^*)(\overline{h})$, and introduce the seminorm
\begin{equation}
p(a(h))=\sup_{\pi\in\rep{\frak W}}\left\|\int_\r dt\ h(t)
  \pi\left(\alpha_t^\ww (a)\right)\right\|,
\label{1.1.1}
\end{equation}
where the supremum extends over all representations of $\ww$. 
The integral on the r.h.s. of \fer{1.1.1} is understood in the weak sense
($t\mapsto \pi(\alpha_t^\ww(a))$ is weakly measurable for any
$\pi\in\rep\ww$), and the norm is the one of 
operators acting on the representation Hilbert space. It is not difficult to
verify that 
\begin{equation*}
{\frak N}=\left\{ a\in{\frak P}\ |\ p(a)=0\right\}
\end{equation*}
is a two-sided $*$ideal in $\frak P$. We can therefore build the quotient
$*$algebra ${\frak P}/{\frak N}$ consisting of equivalence classes $[a]=\{a+n\
|\ a\in{\frak P}, n\in{\frak N}\}$, on which $p$ defines a norm
\begin{equation*}
\|\, [a]\, \|=p(a),\ \ [a]\in{\frak P}/{\frak N},
\end{equation*}
having the $C^*$ property
\begin{equation*}
\|\, [a]^*[a]\, \|=\|\, [a]\, \|^2.
\end{equation*}
The $C^*$-algebra ${\frak A}_f$ of the field is defined to be the closure of
the quotient in this norm, 
\begin{equation*}
{\frak A}_f=\overline{{\frak P}/{\frak N}}^{\, \|\cdot\|}.
\end{equation*}
Notice that every $\pi_\ww\in\rep\ww$ induces a representation
$\pi_f\in\rep{\frak A}_f$ 
according to $\pi_f(a(h))=\int dt\, h(t)\pi_\ww(\alpha_t^\ww(a))$. The algebra
${\frak A}_f$ can be viewed as a time-averaged version of the Weyl algebra. The
advantage of ${\frak A}_f$ over $\ww$ is that the free field dynamics on
${\frak A}$, defined by 
\begin{equation}
\alpha_t^f(a(h))=a(h_t),\ \ \ h_t(x)=h(x-t),
\label{ffd}
\end{equation}
is a norm-continous $*$automorphism group, i.e.,
$\|\alpha_t^f(a)-a\|\rightarrow 0$, as $t\rightarrow 0$, for any $a\in{\frak
  A}_f$. \\
\indent
There is a one-to-one correspondence between $(\beta,\alpha_t^\ww)$-KMS states
$\omega_\beta^\ww$ on $\ww$ and $(\beta,\alpha_t^f)$-KMS states
$\omega_\beta^f$ on ${\frak A}_f$, given by the relation
\begin{equation*}
\omega_\beta^f(a_1(f_1)\cdots a_n(f_n)) =\int dt_1\cdots dt_n\ f_1(t_1)\cdots f_n(t_n)\ 
\omega_\beta^\ww\left( \alpha_{t_1}^\ww(a_1)\cdots
  \alpha_{t_n}^\ww(a_n)\right).
\end{equation*}
If $(\h, \pi^\beta_\ww,\Omega)$ is the GNS representation of
$(\ww,\omega_\beta^\ww)$ then the one of $({\frak A}_f,\omega_\beta^f)$ is
given by $(\h,\pi^\beta_f,\Omega)$, where 
\begin{eqnarray}
\lefteqn{
\pi^\beta_f\left( a_1(f_1)\cdots a_n(f_n)\right) }\nonumber\\
&&
=\int dt_1\cdots dt_n\ f_1(t_1)\cdots f_n(t_n) \pi^\beta_\ww\left( \alpha_{t_1}^\ww(a_0)\cdots
  \alpha_{t_n}^\ww(a_n)\right).
\label{1.1.2}
\end{eqnarray}
It follows that any unitary group implementing the free dynamics relative to
$\pi^\beta_\ww$  implements it in the representation $\pi^\beta_f$, and
conversely.

\subsubsection{The algebra $\frak A$ and the regularized dynamics
  $\alphaepsilon$} 
The $C^*$-algebra $\frak A$ describing the ``observables'' of the combined
system is the tensor product algebra
\begin{equation}
{\frak A}={\frak A}_p\otimes{\frak A}_f.
\label{algobs}
\end{equation}
Here, ${\frak A}_p={\cal B}(\h_p)$ is the $C^*$-algebra of all bounded
operators on the particle Hilbert space 
\begin{equation}
\h_p=\cx\oplus L^2(\r_+,de;\hh)\equiv\cx\oplus\dirint_{\r_+}\hh_e\ de,
\label{a16}
\end{equation}
where  $de$ is the Lebesgue measure on $\r_+$, $\hh$ is a (separable) Hilbert
space, and 
the r.h.s. is the constant fibre direct 
integral with $\hh_e\cong\hh$, $e\in \r_+$. An element in $\h_p$ is given by
$\psi=\{\psi(e)\}_{e\in \{E\}\cup\r_+}$, where $\psi(E)\in\cx$, and
$\psi(e)\in\hh$, $e\in\r_+$. $\h_p$ is a Hilbert space with inner product 
\begin{equation*}
\scalprod{\psi}{\phi}=\overline{\psi(E)}\phi(E)+\int_{\r_+}\scalprod{\psi(e)}{\phi(e)}_{\hh}\
de.
\end{equation*}

Let $\alpha_t^p$ denote the $*$automorphism group on ${\frak A}_p$ given by
\begin{equation*}
\alpha_t^p(A)= e^{it H_p}Ae^{-itH_p},
\end{equation*}
where $H_p$ is a selfadjoint operator on $\h_p$, which is diagonalized  
 by the direct
integral decomposition of $\h_p$: 
\begin{equation}
H_p=E\oplus\dirint_{\r_+}e\ de, \mbox{\ \ \ \ \ for some $E<0$.}
\label{a17}
\end{equation}
The domain of definition of $H_p$ is given by 
\begin{equation}
\dom(H_p)=\cx\oplus \left\{\left.\psi\in\dirint_{\r_+}\hh_e de\ \right| \
  \int_{\r_+}e^2\|\psi(e)\|^2_{\hh} de<\infty\right\}.
\label{a18}
\end{equation}

The dense set $C_0^\infty(\r_+;\hh)\equiv C_0^\infty$ consists of all elements
$\psi\in\h_p$ 
s.t. the support, $\supp(\psi\upharpoonright{\r_+})$,  
is a compact set in the open half-axis $(0,\infty)$, and s.t. $\psi$ is
infinitely many times 
continuously differentiable as an $\hh$-valued function. Clearly,
$C_0^\infty\subset\dom(H_p)$, and 
since $e^{itH_p}$ leaves $C_0^\infty$ invariant, it follows that
$C_0^\infty$ is a core for $H_p$. It is sometimes practical to identify
$\cx\cong \cx\varphi_0$, and we say that $\varphi_0$ is the eigenfunction of
$H_p$ corresponding to the eigenvalue $E$. \\

{\it Example.\ }
This model is inspired by considering a block-diagonal Hamiltonian $H_p$ on
the Hilbert space ${\mathbb C}\oplus L^2(\r^3,d^3x)$, with 
$H_p\upharpoonright\cx=E<0$, $H_p\upharpoonright
L^2(\r^3,d^3x)=-\Delta$. Passing to a diagonal representation of the Laplacian
(Fourier transform), we have the
following identifications, using polar coordinates:
\begin{eqnarray*}
\h_p&=&\cx\oplus L^2(\r^3,d^3k)\\
&=&\cx\oplus L^2(\r_+\times S^2,|k|^2d|k|\times d\Sigma)\\
&=&\cx\oplus L^2(\r_+,|k|^2d|k|; L^2(S^2,d\Sigma))\\
&=&\cx\oplus L^2(\r_+,d\mu;\hh),
\end{eqnarray*}
where we set $\hh=L^2(S^2,d\Sigma)$, and make the change of
variables $|k|^2=e$, so that $d\mu(e)=\mu(e)de$, with
$\mu(e)=\frac{1}{2}\sqrt{e}$. To arrive at the form \fer{a16}, \fer{a17}  of
$\h_p$, $H_p$, 
we use the unitary map $U:
L^2(\r_+,d\mu;\hh)\rightarrow L^2(\r_+,de;\hh)$, given by
\begin{equation*}
\psi\mapsto U\psi=\sqrt{\mu}\psi.
\end{equation*}
If $H_p$ is the operator of multiplication by $e$ on $L^2(\r_+,d\mu;\hh)$,
then its transform, $UH_pU^{-1}$, is the operator of multiplication by $e$ on
$L^2(\r_+,de;\hh)$. \\

We define the non-interacting time-translation $*$automorphism group of $\frak
A$ (the free dynamics)  by
\begin{equation*}
\alpha_{t,0}:=\alpha_t^p\otimes\alpha_t^f.
\end{equation*}
Given $\epsilon\neq 0$, set
\begin{equation}
V^{(\epsilon)}:=\sum_\alpha G_\alpha\otimes \frac{1}{2i\epsilon}\left\{(W(\epsilon g_\alpha))(h_\epsilon)- (W(\epsilon
  g_\alpha))(h_\epsilon)^*\right\} \in {\frak A},
\label{vepsilon}
\end{equation}
where the sum is over finitely many indices $\alpha$, with 
$G_\alpha=G_\alpha^*\in{\cal B}(\h_p)$, $g_\alpha\in L_0^2$, for all
$\alpha$, and 
where $h_\epsilon$ is an approximation of the Dirac distribution localized at
zero. To be specific, we can take
$h_\epsilon(t)=\frac{1}{\epsilon}e^{-t^2/\epsilon^2}$. 
For any value of the real coupling constant $\lambda$, the norm-convergent
Dyson series    
\begin{eqnarray}
\lefteqn{
\alpha_{t,0}(A)}\nonumber\\
&+&\sum_{n\geq 1}(i\lambda)^n\int_0^t
dt_1\cdots\int_0^{t_{n-1}}dt_n \left[
  \alpha_{t_n,0}(V^{(\epsilon)}),\left[\cdots\left[
      \alpha_{t_1,0}(V^{(\epsilon)}),
      \alpha_{t,0}(A)\right]\cdots\right]\right]\nonumber\\
&=:&\alphaepsilon(A),
\label{alphaepsilon}
\end{eqnarray}
where $A\in\frak A$, 
defines a $*$automorphism group on $\frak A$. The multiple integral in
\fer{alphaepsilon} is understood in the product topology coming from the
strong topology of ${\cal B}(\h_p)$ and the norm topology of ${\frak A}_f$.

One should view $\alphaepsilon$
as a {\it regularized 
  dynamics}, in the sense that it has a limit, as $\epsilon\rightarrow 0$, in
suitably chosen representations of $\frak A$; (this is shown below). 

 The functions $g_\alpha\in L^2_0$ are called
{\it form factors}. Using spherical coordinates in $\r^3$, we often write
$g_\alpha=g_\alpha(\omega, \Sigma)$, where $(\omega,\Sigma)\in \r_+\times
S^2$. \\
\indent
In accordance with the direct integral decomposition of $H_p$, the operators
$G_\alpha$ are determined by integral kernels. For $\psi=\{\psi(e)\}\in \h_p$,
 we set 
\begin{equation}
(G_\alpha\psi)(e)=\left\{
\begin{array}{ll}
G_\alpha(E,E)\psi(E)+\displaystyle \int_{\r_+}G_\alpha(E,e')\psi(e') de', &
\mbox{if $e=E$},\\ 
G_\alpha(e,E)\psi(E)+ \displaystyle \int_{\r_+}G_\alpha(e,e')\psi(e')de',
& \mbox{if $e\in\r_+$}.  
\end{array}
\label{opsala}
\right.
\end{equation}
The families of bounded operators
$G_\alpha(e,e'): 
\hh_{e'}\rightarrow\hh_{e}$, with $\hh_E=\cx$,  have the
following  
symmetry properties (guaranteeing that $G_\alpha$ is selfadjoint):
\begin{eqnarray*}
G_\alpha (E,E)&\in&\r,\\
G_\alpha(E,e)^*&=& G_\alpha(e,E),\ \ \forall e\in\r_+,\\
G_\alpha(e,e')^*&=& G_\alpha(e',e),\ \ \forall e,e'\in\r_+.
\end{eqnarray*}
Here, $^*$ indicates taking the adjoint of an operator in ${\cal B}(\hh,\cx)$
or ${\cal B}(\hh)$. \\
\indent
{\it Remarks.\ } 1)\ The map $G_\alpha(E,e):\hh_e\rightarrow \cx$ is
identified (Riesz) with an element $\Gamma_\alpha(e)\in\hh_e$, so that
$G_\alpha(E,e)\psi(e)=\scalprod{\Gamma_\alpha(e)}{\psi(e)}_{\hh_e}$. Then
$G_\alpha(E,e)^*:\cx\rightarrow\hh_e$ is given by
$G_\alpha(E,e)^*z=z\Gamma_\alpha(e)$, for all $z\in\cx$. Consequently, the
above symmetry condition implies that $G_\alpha(e,E)z=z\Gamma_\alpha(e)$. \\
\indent
2)\ Assuming the strong derivatives w.r.t. the two arguments $(e,e')\in\r_+^2$
of
$G_\alpha(\cdot,\cdot)$ exist, we have that 
$\partial_{1,2}G_\alpha(e,e')$ are operators $\hh\rightarrow \hh$. Similarly,
one introduces 
higher derivatives. We assume that all derivatives occuring are bounded
operators on $\hh$. For
$G_\alpha(\cdot,\cdot)\in C^n(\r_+\times\r_+,{\cal B}(\hh))$, it is easily verified that the above
symmetry conditions imply that 
\begin{equation}
\left(\partial_1^{n_1}\partial_2^{n_2}G_\alpha(e,e')\right)^*=\partial_1^{n_2}\partial_2^{n_1}G_\alpha(e',e),
\label{a56}
\end{equation}
for any $n_{1,2}\geq 0$, $n_1+n_2\leq n$,  where $^*$ is the adjoint on
${\cal B}(\hh)$. Similar statements hold for $G_\alpha(E,e)$,
$G_\alpha(e,E)$. \\

The interaction is required to satisfy the following three conditions:
\begin{itemize}
\item[(A1)]  Infrared and ultraviolet behaviour of the form factors: for any
  fixed 
  $\Sigma\in S^2$, $g_\alpha(\cdot,\Sigma)\in C^4(\r_+)$, and there are two constants
  $0<k_1, k_2<\infty$, s.t. if $\omega< k_1$, then 
\begin{equation}
|\partial_\omega^j g_\alpha(\omega,\Sigma)|< k_2 \omega^{p-j}, \mbox{\ \ \ for
 some $p>2$},
\label{IR}
\end{equation}
uniformly in $\alpha$, $j=0,\ldots,4$ and $\Sigma\in S^2$. Similarly, there
are two constants $0<K_1, K_2<\infty$, s.t. if $\omega>K_1$, then
\begin{equation}
|\partial_\omega^j g_\alpha(\omega,\Sigma)|< K_2 \omega^{-q-j}, \mbox{\ \ \ for
 some $q>7/2$}.
\label{UV}
\end{equation}

\item[(A2)]
The map $(e,e')\mapsto G_\alpha(e,e')$ is $C^3(\r_+\times \r_+,{\cal
  B}(\hh))$, and we have  
\begin{eqnarray} 
\int_{\r_+}de\left\| e^{-m_1}\partial_1^{m_2}
  G_\alpha(e,E)\right\|_\hh^2<\infty, 
\label{regul1}\\
\int_{\r_+}de \int_{\r_+} de' \left\| e^{-m_1}(e')^{-m_1'}
  \partial_1^{m_2} \partial_2^{m'_2} G_\alpha(e,e')\right\|_{{\cal B}(\hh)}^2<\infty,
\label{regul2}
\end{eqnarray}
for all integers $m_{1,2}, m'_{1,2}\geq 0$,
s.t. $m_1+m'_1+m_2+m'_2=0,1,2,3$. Moreover,
\begin{equation}
\int_{\r_+}de \int_{\r_+} de'\left\| e \,G_\alpha(e,e')\right\|^2_{{\cal
    B}(\hh)}<\infty.
\label{newcond}
\end{equation}

\item[(A3)] The Fermi Golden Rule Condition. Define a family of bounded
  operators on $\h_p$ by
\begin{equation}
F(\omega,\Sigma)=\sum_\alpha g_\alpha(\omega,\Sigma) G_\alpha.
\label{ef}
\end{equation}
There is an $\epsilon_0>0$, s.t. for $0<\epsilon<\epsilon_0$, we have that
\begin{equation}
\int_{-E}^\infty d\omega\int_{S^2} d\Sigma \ \frac{\omega^2}{e^{\beta\omega}-1}
    p_0F(\omega,\Sigma)\frac{\overline{p}_0\ 
    \epsilon}{(H_p-E-\omega)^2+\epsilon^2} F(\omega,\Sigma)^*p_0\geq
    \gamma p_0,
\label{FGRC}
\end{equation}
for some strictly positive constant $\gamma>0$. Here, $p_0$ is the orthogonal
projection onto the eigenspace $\cx$ of $H_p$ (see \fer{a16}, \fer{a17}), and
$\pbar_0=\bbbone-p_0$ is the projection onto $L^2(\r_+,de;\hh)$. 
\end{itemize}

{\it Remarks.\ } 1) Since $E<0$ we have that $\gamma\sim e^{\beta E}$ decays
exponentially in $\beta$, for large $\beta$.\\
2)  Recalling that $G_\alpha(E,e)$ is identified with
$\Gamma_\alpha(e)\in\hh_e$, see Remark 1) after \fer{opsala} above, we can
rewrite the l.h.s. of \fer{FGRC} as 
\begin{eqnarray*}
\lefteqn{
  \int_{(-E,\infty)\times S^2} d \omega\  d\Sigma 
\int_{\r_+} de\ 
  \frac{\omega^2}{e^{\beta\omega}-1} \frac{\epsilon}{(e-E-\omega)^2+\epsilon^2}  }\\
&&\ \ \ \ \ \ \   \ \ \ \times \sum_{\alpha,\alpha'}\overline{g}_\alpha(\omega,\Sigma)
  \scalprod{\Gamma_\alpha(e)}{\Gamma_{\alpha'}(e)}_\hh 
  g_{\alpha'}(\omega,\Sigma),
\end{eqnarray*}
and this expression has the limit
\begin{equation*}
\int_{(-E,\infty)\times S^2}d\omega \ d\Sigma \ \ \frac{\omega^2}{e^{\beta \omega}-1}\sum_{\alpha,\alpha'} 
\overline{g}_\alpha(\omega,\Sigma)
\scalprod{\Gamma_\alpha(E+\omega)}{\Gamma_{\alpha'}(E+\omega) }_\hh
g_{\alpha'}(\omega,\Sigma),
\end{equation*}
as $\epsilon\rightarrow 0$, because $\Gamma_\alpha(e)$ is
continuous in $e$. 
Consequently, \fer{FGRC} is satisfied if this integral is strictly
positive.

\subsubsection{The reference state $\oref$}

Let $\rho_0$ be a strictly positive density matrix on $\h_p$, i.e., $\rho_0>0$,
${\rm tr}\rho_0=1$, 
and denote by $\omega^p_{\rho_0}$ the state on ${\frak A}_p$ given by
$A\mapsto {\rm tr}\rho_0A$. Let $\omega_\beta^f$ be the
$(\alpha_t^f,\beta)$-KMS state on ${\frak A}_f$ and define the reference state
\begin{equation*}
\oref=\omega^p_{\rho_0}\otimes \omega_\beta^f.
\end{equation*}
The GNS representation $(\h,\pi_\beta,\Oref)$ corresponding to $({\frak A},
\oref)$ is 
explicitly known. It has first been described in [AW]; (we follow [JP] in its
presentation). The representation Hilbert space is 
\begin{equation}
\h=\h_p\otimes\h_p\otimes{\cal F},
\label{7}
\end{equation}
where ${\cal F}$ is a shorthand for the Fock space
\begin{equation}
{\cal F}={\cal F}\left((L^2(\r\times S^2,\ du\times d\Sigma)\right),
\label{calf}
\end{equation}
$du$ being the Lebesgue measure on $\r$, and $d\Sigma$ the uniform measure on
$S^2$. ${\cal F}(X)$ denotes the bosonic Fock space over a (normed vector)
space $X$:
\begin{equation}
{\cal F}(X):=\cx\oplus\bigoplus_{n\geq 1} \left({\cal S} X^{\otimes n}\right),
\label{fockspace}
\end{equation}
where ${\cal S}$ is the projection onto the symmetric subspace of the
tensor product. We adopt standard notation, e.g. $\Omega$ is the vacuum
vector, $[\psi]_n$ is the $n$-particle component of $\psi\in{\cal F}(X)$,
$\d\Gamma(A)$ is the second quantization of the operator $A$ on $X$,
$N=\d\Gamma(\bbbone)$ is the number operator. \\
\indent
The representation map $\pi_\beta:{\frak A}\rightarrow {\cal B}(\h)$ is the
product 
\begin{equation*}
\pi_\beta=\pi_p\otimes\pi^\beta_f,
\end{equation*}
where the $*$homomorphism $\pi_p:{\frak A}_p\rightarrow{\cal
  B}(\h_p\otimes\h_p)$ is given by
\begin{equation*}
\pi_p(A)= A\otimes\bbbone_p.
\end{equation*}
The representation map $\pi_f^\beta:{\frak A}_f\rightarrow{\cal B}({\cal F})$
is determined by the representation map of the Weyl algebra,
$\pi_\ww^\beta:\ww\rightarrow {\cal B}({\cal F})$, according to
\fer{1.1.2}. To describe $\pi_\ww^\beta$, we point out that  $L^2(\r_+\times
S^2)\oplus L^2(\r_+\times 
S^2)$ is isometrically isomorphic to $L^2(\r\times S^2)$
via the map 
\begin{equation}
(f,g)\mapsto h,\ \ h(u,\Sigma)=\left\{
\begin{array}{ll}
u\ f(u,\Sigma), & u>0,\\
u\ g(-u,\Sigma), & u<0.
\end{array}
\right.
\label{9'}
\end{equation}
The representation map $\pi_\ww^\beta$ is given by
\begin{equation*}
\pi_\ww^\beta=\pi_{\rm Fock}\circ{\cal T}_\beta,
\end{equation*}
where the Bogoliubov transformation ${\cal T}_\beta: {\frak
  W}(L^2_0)\rightarrow {\frak W}(L^2(\r\times 
S^2))$ acts as $W(f)\mapsto W(\tau_\beta f)$, with $\tau_\beta: L^2(\r_+\times
  S^2)\rightarrow L^2(\r\times S^2)$ given by 
\begin{equation}
(\tau_\beta f)(u,\Sigma)=\sqrt{\frac{u}{1-e^{-\beta u}}}\left\{
\begin{array}{ll}
\sqrt{u}\ f(u,\Sigma), & u>0,\\
-\sqrt{-u}\ \overline{f}(-u,\Sigma), & u<0.
\end{array}
\right.
\label{9}
\end{equation}
{\it Remarks.} 1) It is easily verified that $\IM\scalprod{\tau_\beta
  f}{\tau_\beta g}_{L^2(\r\times S^2)}=\IM\scalprod{f}{g}_{L^2(\r_+\times S^2)}$, for all
  $f,g\in L_0^2$, so the CCR \fer{CCR} are preserved under the map
  $\tau_\beta$. \\
\indent
2)\ In the limit $\beta\rightarrow\infty$, the r.h.s. of \fer{9} tends to
\begin{equation*}
\left\{
\begin{array}{ll}
u\ f(u,\Sigma), & u>0,\\
0, & u<0,
\end{array}
\right.
\end{equation*}
which is identified, via \fer{9'}, with $f\in L^2_0$. Thus, ${\cal T}_\beta$
reduces 
to the identity (an imbedding), $\pi_\ww^\beta$ becomes the Fock
representation of ${\frak W}(L^2_0)$, as $\beta\rightarrow\infty$, and we
recover the zero temperature situation. \\
\indent
It is useful to introduce the following notation. For $f\in L^2(\r\times
S^2)$, we define unitary operators, $\What(f)$, on the Hilbert space \fer{7},
by  
\begin{equation*}
\What(f)=e^{i\varphi(f)},\ \ f\in L^2(\r\times S^2),
\end{equation*}
where $\varphi(f)$ is the selfadjoint operator on $\cal F$ given by
\begin{equation}
\varphi(f)= \frac{a^*(f)+a(f)}{\sqrt{2}},
\label{fieldop}
\end{equation}
and $a^*(f)$, $a(f)$ are the creation- and annihilation operators on $\cal F$,
smeared out with $f$. One easily verifies that
\begin{equation*} 
\pi_\ww^\beta(W(f))=\What(\tau_\beta f).   
\end{equation*}

The cyclic GNS vector is given by 
\begin{equation*}
\Oref=\Omega_p^{\rho_0}\otimes\Omega, 
\end{equation*}
where $\Omega$ is the vacuum in $\cal F$, and 
\begin{equation}
\Omega_p^{\rho_0}=\sum_{n\geq 0} k_n\varphi_n\otimes\cc_p\varphi_n\in
\h_p\otimes\h_p.
\label{omegapart}
\end{equation}
Here, $\{k^2_n\}_{n=0}^\infty$ is the spectrum of $\rho_0$,
$\{\varphi_n\}$ is an orthogonal basis of eigenvectors of $\rho_0$, and
$\cc_p$ is an antilinear involution on $\h_p$. The origin of $\cc_p$ lies in
the identification of $l^2(\h_p)$ (Hilbert-Schmitt operators on $\h_p$) with
$\h_p\otimes\h_p$, via $|\varphi\rangle\langle\psi|\mapsto
\varphi\otimes\cc_p\psi$. We fix a convenient choice for $\cc_p$: it is the
antilinear involution on $\h_p$ that has the effect of taking complex
conjugates of components of vectors, in the basis in which the
Hamiltonian $H_p$ is diagonal, i.e.,
\begin{equation*}
(\cc_p\psi)(e)=\left\{
\begin{array}{ll}
\overline{\psi(e)}\in\cx, & e=E,\\
\overline{\psi(e)}\in\hh, & e\in[0,\infty).
\end{array}
\right.
\end{equation*}
By $\overline{\psi(e)}\in\hh$ for $e\in[0,\infty)$, we understand the element
in $\hh$ obtained by complex conjugation of the components of $\psi(e)\in\hh$,
in an arbitrary, but fixed, orthonormal basis of $\hh$. This $\cc_p$ is also
called the time reversal operator, and we have
\begin{equation*}
\cc_p H_p \cc_p=H_p.
\end{equation*}

\subsubsection{The $W^*$-dynamical system $(\mm_\beta,\sigma_{t,\lambda})$}
Let $\mm_\beta$ be the von Neumann algebra obtained by taking the weak closure
(or equivalently, the double commutant) of $\pi_\beta({\frak A})$ in ${\cal
  B}(\h)$:
\begin{equation*}
\mm_\beta={\cal B}(\h_p)\otimes\bbbone_p\otimes\pi^\beta_f({\frak A}_f)''
\subset {\cal B}(\h).
\end{equation*}
Since $\rho_0$ is strictly positive, $\Omega_p^{\rho_0}$ is cyclic and
separating for the von Neumann algebra $\pi_p({\frak A}_p)''={\cal
  B}(\h_p)\otimes\bbbone_p$. Similarly, $\Omega$ is cyclic and separating for
$\pi_f^\beta({\frak A}_f)''$, since it is the GNS vector of a KMS state (see
e.g. [BRII]). Consequently, $\Oref$ is cyclic and separating for
$\mm_\beta$. Let $J$ be the modular conjugation operator associated to
$(\mm_\beta,\Oref)$. It is given by
\begin{equation}
J=J_p\otimes J_f,
\label{J}
\end{equation}
where, for $\varphi,\psi\in\h_p$,
\begin{equation*}
J_p\left(\varphi\otimes {\cal C}_p\psi\right)=\psi\otimes {\cal C}_p\varphi,
\end{equation*}
and, for $\psi=\{[\psi]_n\}_{n\geq 0}\in{\cal F}$,
\begin{eqnarray*}
[J_f\psi]_n(u_1,\ldots,u_n)&=& \overline{[\psi]_n(-u_1,\ldots,-u_n)},\ \
\mbox{for $n\geq 1$},\\
\  [J_f\psi]_0&=&\overline{[J_f\psi]_0}\in {\mathbb C}.
\end{eqnarray*}
Clearly, $J\Oref=\Oref$, and one verifies that
\begin{eqnarray}
J_p\pi_p(A)J_p &=&\bbbone_p\otimes \cc_p A\cc_p,\label{jparticle}\\
J_f\pi_\ww^\beta(W(f))J_f&=& \What(-e^{-\beta
  u/2}\tau_\beta(f))=\What(e^{-\beta 
  u/2}\tau_\beta(f))^*,
\label{jfield}
\end{eqnarray}
for $f\in L^2_0$. More generally, for $f\in L^2(\r\times S^2)$, 
$J_f\What(f)J_f=\What(\overline{f(-u,\Sigma)})$. \\

We now construct a unitary implementation of 
$\alphaepsilon$ w.r.t. $\pi_\beta$.  
Recall that $\pi_\beta =\pi_p\otimes\pi^\beta_f$, where $\pi_p:{\cal
  B}(\h)\rightarrow {\cal B}(\h_p\otimes\h_p)$ is continuous w.r.t. the strong
topologies and $\pi_f^\beta:{\frak A}_f\rightarrow{\cal B}({\cal F})$  is
continuous w.r.t. the norm topologies (because it is a $*$ homomorphism). We
thus have, for $A\in\frak A$,
\begin{eqnarray}
\lefteqn{
  \pi_\beta(\alphaepsilon(A))}\nonumber\\
&=&\pi_\beta(\alpha_{t,0}(A))+\sum_{n\geq 1}(i\lambda)^n\int_0^tdt_1\cdots
  \int_0^{t_{n-1}}dt_n
  \Big[\pi_\beta(\alpha_{t_n,0}(V^{(\epsilon)})),\Big[\cdots\nonumber\\
&& \cdots
  \big[\pi_\beta(\alpha_{t_1,0}(V^{(\epsilon)})),\pi_\beta(\alpha_{t,0}(A))\Big]\cdots
  \Big]\Big].
\label{pidyson}
\end{eqnarray}
Because  
\begin{equation*}
\pi_p(\alpha_t^p(A))=e^{itH_p}Ae^{-itH_p}\otimes\bbbone_p= e^{it(H_p\otimes\bbbone_p
  -\bbbone_p\otimes H_p)}\pi_p(A) e^{-it(H_p\otimes\bbbone_p
  -\bbbone_p\otimes H_p)},
\end{equation*}
and 
\begin{eqnarray*}
\pi_\ww^\beta(\alpha_t^\ww(W(f)))&=&\pi_\ww^\beta(W(e^{i\omega t}f))=
\What(e^{iut}\tau_\beta(f))\\
&=&e^{it\d\Gamma(u)} \What(\tau_\beta(f)) e^{-it\d\Gamma(u)}
=e^{it\d\Gamma(u)}\pi_\ww^\beta(W(f))e^{-it\d\Gamma(u)},
\end{eqnarray*}
so that 
\begin{equation*}
\pi_f^\beta(\alpha_t^f(a))=e^{it\d\Gamma(u)} \pi_f^\beta(a)
e^{-it\d\Gamma(u)},\ \ a\in{\frak A}_f,
\end{equation*}
we find that
\begin{equation*}
 \sigma_{t,0}(\pi_\beta(A)):= \pi_\beta(\alpha_{t,0}(A))= e^{itL_0}\pi_\beta
  (A) e^{-itL_0},
\end{equation*}
for all $A\in\frak A$, where $L_0$ is the selfadjoint operator on $\h$, given
by   
\begin{equation}
L_0=H_p\otimes\bbbone_p-\bbbone_p\otimes H_p +\d\Gamma(u),
\label{L_0}
\end{equation}
commonly called the (non-interacting, standard) Liouvillian. One easily
verifies that  
\begin{equation}
Je^{itL_0}=e^{itL_0}J. 
\label{jllj}
\end{equation}

{\it Remark.\ } There are other selfadjoint operators generating unitary
implementations of 
$\sigma_{t,0}$ on $\h$. Indeed, we may add to $L_0$ any selfadjoint
operator $L_0'$ affiliated with the commutant ${\mm_\beta}'$; then $L_0+L_0'$ 
 still generates a unitary implementation of $\sigma_{t,0}$ on $\h$. However,
 the additional condition 
 \fer{jllj} fixes $L_0$ uniquely, and the generator
 of this unitary group is called the {\it standard} Liouvillian for
 $\sigma_{t,0}$. This terminology has been used before in [DJP]. The importance of considering the standard Liouvillian (as
 opposed to other generators of the dynamics) lies in the fact that its
 spectrum is related to the dynamical properties of the system; see
 Theorem \ref{nostatstatesthm}.\\

Notice that $\sigma_{t,0}$ is a group of
$*$automorphisms of $\pi_\beta(\frak A)$, in particular,
$e^{itL_0}\pi_\beta({\frak A})e^{-itL_0}= \pi_\beta(\frak A)$, $\forall
t\in\r$. From  Tomita-Takesaki theory, we know that $J\mm_\beta
J={\mm_\beta}'$ (the commutant), and since 
\begin{equation*}
\sigma_{t,0}\left(J\pi_\beta(V^{(\epsilon)})J\right)=J\sigma_{t,0}\left(\pi_\beta(V^{(\epsilon)})\right)J=J\pi_\beta(
\alpha_{t,0}(V^{(\epsilon)}))J\in{\mm_\beta}', 
\end{equation*}
we can write the multicommutator in \fer{pidyson} as
\begin{eqnarray*}
\Big[\sigma_{t_n,0}\left(\pi_\beta(V^{(\epsilon)})-J\pi_\beta(V^{(\epsilon)})J\right),\Big[\cdots\\
\cdots\Big[
  \sigma_{t_1,0}\left(\pi_\beta(V^{(\epsilon)})-J\pi_\beta(V^{(\epsilon)})J\right),\sigma_{t,0}(\pi_\beta(A))\Big]\cdots\Big]\Big].
\end{eqnarray*}
It follows that the r.h.s. of \fer{pidyson} defines a $*$automorphism group of
$\pi_\beta(\frak A)$, $\sigmaepsilon$, which is implemented unitarily by
\begin{equation*}
\sigmaepsilon(\pi_\beta(A))= \pi_\beta(\alphaepsilon(A))=
e^{itL_\lambda^{(\epsilon)}} \pi_\beta(A) 
e^{-itL_\lambda^{(\epsilon)}},
\end{equation*}
with
\begin{equation*}
L_\lambda^{(\epsilon)}= L_0 +\lambda\pi_\beta(V^{(\epsilon)}) -\lambda
J\pi_\beta(V^{(\epsilon)}) J.
\end{equation*}
It is not difficult to see (using Theorem \ref{nelsonthm}) that the {\it
  regularized 
  Liouvillian $L_\lambda^{(\epsilon)}$} is essentially selfadjoint on 
\begin{equation*}
\dom=C_0^\infty\otimes C_0^\infty\otimes ({\cal F}(C_0^\infty(\r\times
S^2))\cap {\cal F}_0)\subset \h, 
\end{equation*}
where ${\cal F}_0$ is the finite-particle subspace. Moreover, we have that
$Je^{itL_\lambda^{(\epsilon)}}=e^{itL_\lambda^{(\epsilon)}}J$. We now explain
how to remove 
the regularization ($\epsilon\rightarrow 0$), obtaining a weak$*$ continuous
$*$automorphism group $\sigma_{t,\lambda}$ of the von Neumann algebra
$\mm_\beta$. We recall that a $*$automorphism group $\tau_t$ on a von Neumann algebra $\frak M$ is called {\it weak* continuous} iff $t\mapsto \omega(\tau_t(A))$ is continuous, for all $A\in \frak M$ and for all normal states $\omega$ on $\frak M$. From 
\begin{eqnarray*}
\pi_\beta(V^{(\epsilon)})&=&\sum_\alpha
G_\alpha\otimes\bbbone_p\otimes\frac{1}{2i\epsilon} \int_\r dt\ h_\epsilon(t)
\left\{
  \widehat{W}(e^{iut}\epsilon\tau_\beta(g_\alpha)) -\right.\\
&&\ \ \ \ \ \ \ \left.
  \widehat{W}(e^{iut}\epsilon\tau_\beta(g_\alpha))^*\right\}\\
J\pi_\beta(V^{(\epsilon)})J&=& \sum_\alpha \bbbone_p\otimes\cc_p G_\alpha\cc_p\otimes
\frac{1}{2i\epsilon} \int_\r dt\ h_\epsilon(t) \left\{
  \widehat{W}(e^{iut}\epsilon e^{-\beta u/2}\tau_\beta(g_\alpha)) -\right.\\
&&\ \ \ \ \ \ \ \left.
  \widehat{W}(e^{iut}\epsilon e^{-\beta u/2}\tau_\beta(g_\alpha))^*\right\},
\end{eqnarray*}
where we recall that $h_\epsilon(t)=\frac{1}{\epsilon} e^{-t^2/\epsilon^2}$
approximates the Dirac delta distribution concentrated at zero, 
one verifies that, in the strong sense on $\dom$, 
\begin{eqnarray*}
\lim_{\epsilon\rightarrow 0}\pi_\beta (V^{(\epsilon)}) &=&\sum_\alpha
G_\alpha\otimes \bbbone_p\otimes \varphi (\tau_\beta(g_\alpha)),\\
\lim_{\epsilon\rightarrow 0}J\pi_\beta (V^{(\epsilon)})J &=&\sum_\alpha
\bbbone_p\otimes \cc_p G_\alpha\cc_p\otimes \varphi (e^{-\beta
  u/2}\tau_\beta(g_\alpha)),
\end{eqnarray*}
where the operator $\varphi(f)$ has been defined in \fer{fieldop}. 
The symmetric operator $L_\lambda$, defined on $\dom$ by 
\begin{equation}
L_\lambda= L_0 +\lambda I,\label{L}
\end{equation}
with
\begin{equation}
I= \sum_\alpha G_\alpha\otimes\bbbone_p\otimes\varphi(\tau_\beta(g_\alpha))
-\bbbone_p\otimes \cc_pG_\alpha\cc_p\otimes\varphi (e^{-\beta
  u/2}\tau_\beta(g_\alpha)),
\label{I}
\end{equation}
is essentially selfadjoint on $\dom$, for any real value of $\lambda$; (this
will be shown to be a consequence of  Theorem 
\ref{nelsonthm}). Using Theorem \ref{IDthm} on invariance of domains, the
Duhamel formula gives 
\begin{equation*}
e^{itL_\lambda^{(\epsilon)}}=e^{itL_\lambda}-i\lambda\int_0^t e^{isL_\lambda}
\left(I-\pi_\beta(V^{(\epsilon)})+J\pi_\beta(V^{(\epsilon)})J\right)
e^{-i(s-t)L_\lambda^{(\epsilon)}}
\end{equation*}
as operators defined on $\dom$, 
from which it follows that $e^{itL_\lambda^{(\epsilon)}}\rightarrow e^{itL_\lambda}$, as
$\epsilon\rightarrow 0$, in the strong sense on $\h$. Consequently, for
$A\in\pi_\beta(\frak A)$, we have
$\sigmaepsilon(A)\rightarrow\sigma_{t,\lambda}(A)$, in the $\sigma$-weak
topology of ${\cal B}(\h)$. 
Notice that for $A\in\pi_\beta(\frak A)$, we have
  $\sigma_{t,\lambda}(A)\in\mm_\beta$, because
  $\sigma_{t,\lambda}(A)=\mbox{w-}\lim_{\epsilon\rightarrow
  0}\sigmaepsilon(A)$, $\sigmaepsilon(A)\in\pi_\beta({\frak
  A})\subset \mm_\beta$, and $\mm_\beta$ is weakly closed.  
Clearly, $\sigma_{t,\lambda}$ is a $\sigma$-weakly
continuous $*$automorphism group of ${\cal B}(\h)$. If $A\in\mm_\beta$, 
there is a net $\{A_\alpha\}_{\alpha\in{\cal I}}\subset\pi_\beta(\frak A)$,
s.t. $A_\alpha\rightarrow A$, in the weak operator topology. Thus, since
$\sigma_{t,\lambda}$ is weakly continuous, we conclude that 
\begin{equation*}
\sigma_{t,\lambda}(A)=\mbox{w-}\lim_{\alpha}\sigma_{t,\lambda}
(A_\alpha)\in\mm_\beta.
\end{equation*}
We summarize these considerations in a proposition. 

\begin{proposition} 
\stepcounter{theorem}
$(\mm_\beta,\sigma_{t,\lambda})$ is a $W^*$-dynamical
  system, i.e. $\sigma_{t,\lambda}$ is a weak$*$ continuous group of
  $*$automorphisms of the von Neumann algebra $\mm_\beta$. Moreover,
  $\sigma_{t,\lambda}$ is unitarily implemented by $e^{itL_\lambda}$, where $L_\lambda$ is
  given in \fer{L}, \fer{I}, and 
\begin{equation*}
Je^{itL_\lambda}=e^{itL_\lambda}J, \mbox{\ \ for all $t\in\r$}. 
\end{equation*}
\end{proposition}

\subsubsection{Kernel of $L_\lambda$ and normal invariant states}

Let $\pp$ be the natural cone associated with $(\mm_\beta,\Oref)$, i.e., $\pp$
is the norm closure of the set
\begin{equation*}
\{ AJA\Oref\ |\ A\in\mm_\beta\}\subset \h.
\end{equation*}
The data $(\mm_\beta,\h,J,\pp)$ is called the {\it standard form} of the von
Neumann algebra $\mm_\beta$. We have constructed $J$ and $\pp$ explicitly,
starting from the cyclic and separating vector
$\Oref$. There is, however, a general theory of standard forms of von Neumann
algebras; see [BRI], [Ara], [Con] for the case of $\sigma$-finite von Neumann
algebras (as in our case), or [Haa] for the general case. Among the
properties of standard forms, we mention here only the following:
\begin{itemize}
\item[(P)] For every normal state $\omega$ on $\mm_\beta$, there exists a
  unique 
  $\xi\in\pp$, s.t. $\omega(A)=\scalprod{\xi}{A\xi}$, $\forall A\in
  \mm_\beta$. 
\end{itemize}
Recall that a state $\omega$ on $\mm_\beta\subset{\cal B}(\h)$ is called
normal iff it is $\sigma$-weakly continuous, or, equivalently, iff it is given
by  a density matrix $\rho\in l^1(\h)$,
as $\omega(A)=\mbox{tr}\rho A$, for all $A\in\mm_\beta$. 
The uniqueness of the representing vector in the natural cone, according to
(P), allows us to establish the following connection between the kernel of $L_\lambda$
and the normal invariant states (see also e.g. [DJP]).
\begin{theorem}
\label{nostatstatesthm}
\stepcounter{proposition}
If $L_\lambda$ does not have a zero eigenvalue, i.e., if $\mbox{\rm ker}L_\lambda=\{0\}$, then
there does not exist
any $\sigma_{t,\lambda}$-invariant normal state on $\mm_\beta$.
\end{theorem}

{\it Proof.\ } We show below that, for all $t\in\r$, 
\begin{equation}
e^{itL_\lambda}\pp=\pp.
\label{pinv}
\end{equation}
If  $\omega$ is a normal state on $\mm_\beta$, invariant under
$\sigma_{t,\lambda}$, i.e., such that $\omega\circ\sigma_{t,\lambda}=\omega$,
for all $t\in\r$, then, for a unique $\xi\in\pp$, 
\begin{equation*}
\omega(A)=\scalprod{\xi}{A\xi}=\omega(\sigma_{t,\lambda}(A))=\scalprod{
  e^{-itL_\lambda}\xi}{A e^{-itL_\lambda}\xi}.
\end{equation*}
Since \fer{pinv} holds, and due to the uniqueness of the vector in $\pp$
representing a given state, we
conclude that $e^{itL_\lambda}\xi=\xi$, for all $t\in\r$, i.e. $L_\lambda$ has a zero eigenvalue
with eigenvector $\xi$.\\
\indent
 We now show \fer{pinv}. Notice that \fer{pinv} is
equivalent to $e^{itL_\lambda}\pp\subseteq\pp$. Since $\pp$ is a closed set, it is
enough to show that for all $A\in\mm_\beta$, 
$e^{itL_\lambda}AJA\Oref\in\pp$. Since $e^{itL_\lambda}J=Je^{itL_\lambda}$,
$e^{itL_\lambda}Ae^{-itL_\lambda}\in\mm_\beta$, for all $A\in\mm_\beta$, and
$BJBJ\pp\subset\pp$, for all $B\in\mm_\beta$, we only need to prove that 
\begin{equation}
e^{itL_\lambda}\Oref\in\pp.
\label{pinv2}
\end{equation}
The Trotter product formula gives
\begin{equation*}
e^{itL_\lambda}\Oref=\lim_{n\rightarrow\infty}\left( e^{i\frac{t}{n}\lambda I}
  e^{i\frac{t}{n}L_0}\right)^n \Oref,
\end{equation*}
and, since $\pp$ is closed, \fer{pinv2} holds provided the general term under
the 
limit is in $\pp$, for all $n\geq 1$. We show that $e^{isL_0}\pp=\pp$ and
$e^{is\lambda I}\pp=\pp$, for all $s\in\r$. Remarking that
\begin{equation*}
e^{isL_0}\Oref=\left(e^{isH_p}\otimes e^{-isH_p}\otimes
  e^{is\d\Gamma(u)}\right) \Oref =
\left(e^{isH_p}\otimes\bbbone_p \right)J
\left(e^{isH_p}\otimes\bbbone_p\right) \Oref,
\end{equation*}
where we use that $J_p (e^{isH_p}\otimes\bbbone_p) J_p=\bbbone_p\otimes \cc_p
e^{isH_p}\cc_p=\bbbone_p\otimes e^{-isH_p}$, recalling that $e^{isL_0}$
implements $\sigma_{t,0}$, and arguing as above, we see that
$e^{itL_0}\pp=\pp$. \indent
The Trotter product formula gives
\begin{eqnarray*}
\lefteqn{
\exp\left\{ is \sum_{\alpha=1}^N
  G_\alpha\otimes\bbbone_p\otimes\varphi(\tau_\beta 
  (g_\alpha))-J  G_\alpha\otimes\bbbone_p\otimes\varphi(\tau_\beta
  (g_\alpha)) J\right\}\xi=}\\
&&\lim_{n_1\rightarrow\infty} \left\{
  \left(e^{i\frac{s}{n_1}G_1}\otimes\bbbone_p\otimes 
  \What\left( \frac{s}{n_1}\tau_\beta(g_\alpha)\right)\right) J\left(
  e^{i\frac{s}{n_1}G_1}\otimes\bbbone_p\otimes 
  \What\left( \frac{s}{n_1}\tau_\beta(g_\alpha)\right)J\right) \right.\\
&&\times \left. \exp\left[i\frac{s}{n_1}\sum_{\alpha=2}^N\left(
      G_\alpha\otimes\bbbone_p\otimes\varphi(\tau_\beta(g_\alpha)) -J 
G_\alpha\otimes\bbbone_p\otimes\varphi(\tau_\beta(g_\alpha))J\right)
\right]\right\}^{n_1}\xi,
\end{eqnarray*}
for all $\xi\in\pp$, 
and we may apply Trotter's formula repeatedly to
conclude that, since $AJAJ\pp\subset\pp$, for $A\in\mm_\beta$, and $\pp$
is closed, we have that $e^{is\lambda I}\pp=\pp$, for all $s\in\r$.\hfill
$\blacksquare$ 

\ \\
{\it Remark. \ } The proof of Theorem \ref{nostatstatesthm} uses property (P),
which is satisfied in our case, because $\Oref$ is 
cyclic and separating for $\mm_\beta$. This, in turn, is true because $\rho_0$
has been chosen to be strictly positive. One may start with any reference
state of 
the form $\omega_\rho^p\otimes\omega_\beta^f$, where $\rho$ is any density
matrix on $\h_p$; it may be of finite rank. The resulting von Neumann algebra
(obtained as the weak closure of $\frak A$ when represented on the GNS Hilbert
space corresponding to $({\frak A},\omega^p_\rho\otimes \omega^f_\beta)$) is
$*$isomorphic to $\mm_\beta$. This is the reason we have not added to
$\mm_\beta$ an index for the density matrix $\rho_0$. More specifically, the GNS
representation of $({\frak A},\omega^p_\rho\otimes \omega^f_\beta)$ is given
by $(\h_1,\pi_1,\Omega_1)$, where 
\begin{eqnarray*}
\h_1&=&\h_p\otimes {\cal K}_\rho\subseteq \h_p\otimes\h_p,\\
\pi_1(A\otimes (W(f))(h))&=& A\otimes\bbbone_p\otimes \int_\r dt\ h(t)
\What(e^{iut}\tau_\beta(f)),\\ 
\Omega_1&=&\Omega_p^\rho\otimes\Omega.
\end{eqnarray*}
Here, ${\cal K}_\rho$ is the closure of $\ran\rho$, $\Omega_p^\rho$ is given
 as in equation \fer{omegapart}. Consequently, 
\begin{equation*}
\pi_1({\frak A})'' ={\cal B}(\h_p)\otimes\bbbone_p\upharpoonright_{{\cal
    K}_\rho}\otimes \pi_f^\beta({\frak A}_f)''\cong \mm_\beta.
\end{equation*}
In particular, $\pi_1({\frak A})''$ and $\mm_\beta$ have {\it the same set of
  normal states}. Thus, our particular choice for the reference state is immaterial when
examining properties of normal states. One may express this in the
following way: 
$(\mm_\beta, \h,J,\pp)$ is a standard form for all the von Neumann algebras
obtained from any reference state $({\frak A},\omega^p_\rho\otimes
\omega^f_\beta)$.

\subsection{Result on thermal ionization}
\label{tisect}
Our main result in this paper is that the $W^*$-dynamical
system $(\mm_\beta,\sigma_{t,\lambda})$ introduced above does not have any
normal invariant states.

\begin{theorem}
\label{mainthm}
\stepcounter{proposition}
Assume conditions (A1)-(A3) hold. For any inverse temperature
$0<\beta<\infty$ there is a constant, $\lambda_0(\beta)>0$, 
proportional to $\gamma$ given in \fer{FGRC},  such that the
following holds. If 
$0<|\lambda|<\lambda_0$ then the Liouvillian $L_\lambda$ given in \fer{L} and
\fer{I} does not have any eigenvalues. 
\end{theorem}

{\it Remark.\ } Since $\gamma$ decays exponentially in $\beta$, for large
$\beta$, Theorem \ref{mainthm} is a high temperature result ($\beta$ has to be
small for reasonable values of the coupling constant $\lambda$). From 
physics it is clear that 
thermal ionization takes place for {\it arbitrary positive temperatures} (but
not at zero temperature, where the coupled system has a ground state). \\ 
\indent
Combining Theorems \ref{mainthm} and \ref{nostatstatesthm} yields our main
result about thermal ionization.

\begin{theorem}{\bf (Thermal ionization)\ }
\label{2thm}
\stepcounter{proposition}
Under the assumptions of Theorem \ref{mainthm}, there do not exist any normal
$\sigma_{t,\lambda}$-invariant states on $\mm_\beta$. 
\end{theorem}

{\it Remark.\ } For $\lambda=0$, the state $\omega_0$, determined by
the vector $\Omega_p^0\otimes\Omega$, where
$\Omega_p^0=\varphi_0\otimes\varphi_0\in\h_p\otimes\h_p$, and $\varphi_0$
is the eigenvector of $H_p$, is a normal $\sigma_{t,0}$-invariant state on
$\mm_\beta$. 
As we have explained in the introduction, the physical interpretation of
Theorem \ref{2thm} is that a single atom coupled to black-body radiation at a
sufficiently high positive temperature will always end up being
ionized.\\
\indent
The proof of Theorem \ref{mainthm} is based on a novel virial theorem.

\section{Virial theorems and the positive commutator method}
\label{vtsect}

\subsection{Two abstract virial theorems}
\label{absvthm}

Let $\h$ be a Hilbert space, $\dom\subset\h$ a core for a selfadjoint
operator $Y\geq\bbbone$, and $X$ a symmetric operator on 
$\dom$. We say the triple $(X,Y,\dom)$ satisfies the {\it GJN
  (Glimm-Jaffe-Nelson) 
  Condition}, or that $(X,Y,\dom)$ is a {\it GJN-triple}, if there is a
constant $k<\infty$, s.t. for all $\psi\in\dom$: 
\begin{eqnarray}
\|X\psi\|&\leq& k\|Y\psi\| \label{nc1}\\
\pm i\left\{\scalprod{X\psi}{Y\psi}-\scalprod{Y\psi}{X\psi}\right\}&\leq&
k\scalprod{\psi}{Y\psi}.
\label{nc2}
\end{eqnarray}
Notice that if $(X_1,Y,\dom)$ and $(X_2,Y,\dom)$ are GJN triples, then so is
$(X_1+X_2,Y,\dom)$. Since $Y\geq\bbbone$, inequality \fer{nc1} is equivalent
to  
\begin{equation*}
\| X\psi\|\leq k_1\|Y\psi\|+k_2\|\psi\|,
\end{equation*}
for some $k_1, k_2<\infty$.

\begin{theorem}{\bf (GJN commutator theorem) }
\label{nelsonthm}
\stepcounter{proposition}
If $(X,Y,\dom)$ satisfies
  the GJN Condition, then $X$ determines a selfadjoint operator (again
  denoted by $X$), s.t. $\dom(X)\supset\dom(Y)$. Moreover, $X$ is essentially
  selfadjoint on any core for $Y$, and \fer{nc1} is valid for all
  $\psi\in\dom(Y)$.
\end{theorem}

Based on the GJN commutator theorem, we next
describe the setting for a general {\it virial theorem}. Suppose one is given
a selfadjoint operator $\Lambda\geq\bbbone$ with core $\dom\subset\h$, and 
 operators  $L, A, N, D, C_n$, $n=0,1,2,3$, all symmetric on $\dom$, and
 satisfying  
\begin{eqnarray}
\scalprod{\varphi}{D\psi}&=&i\left\{
  \scalprod{L\varphi}{N\psi}-\scalprod{N\varphi}{L\psi}\right\} \label{44}\\
C_0&=& L\nonumber\\
\scalprod{\varphi}{C_n\psi}&=&i\left\{\scalprod{C_{n-1}\varphi}{A\psi}-\scalprod{A\varphi}{C_{n-1}\psi}\right\},\
  \ n=1,2,3,
\label{45}
\end{eqnarray}
where $\varphi, \psi\in\dom$. We assume that
\begin{itemize}
\item[$\bullet$]  $(X,\Lambda,\dom)$ satisfies the GJN Condition, for
$X=L,N,D,C_n$. Consequently, all these operators determine selfadjoint
operators, which we denote by the same letters.
\item[$\bullet$] $A$ is selfadjoint, $\dom\subset\dom(A)$, and 
$e^{itA}$ leaves $\dom(\Lambda)$
invariant.
\end{itemize}
{\it Remarks.\ } 
1) From the invariance condition $e^{itA}\dom(\Lambda)\subset \dom(\Lambda)$, it follows that for some $0\leq k,k'<\infty$, and all $\psi\in\dom(\Lambda)$,
\begin{equation}
\|\Lambda e^{itA}\psi\|\leq ke^{k'|t|}\|\Lambda\psi\|.
\label{alambda}
\end{equation}
A proof of this can be found in [ABG], Propositions 3.2.2 and 3.2.5.\\
2)\ Condition \fer{nc1} is phrased equivalently as ``$X\leq
kY$, in the sense of Kato on $\dom$''. \\
3)\ One can show that if $(A,\Lambda,\dom)$ satisfies conditions \fer{nc1},
\fer{nc2}, then the above assumption on $A$ holds; see Theorem \ref{IDthm}.

\begin{theorem}{\bf ($1^{\rm st}$ virial theorem) }
\label{virialthm}
\stepcounter{proposition}
Assume that, in addition to \fer{44}, \fer{45}, we have, in the sense of Kato
  on $\dom$, 
\begin{eqnarray}
 D&\leq& kN^{1/2},\label{41}\\
 e^{itA}C_1 e^{-itA}&\leq& ke^{k'|t|}N^p, \mbox{\ \ \ some $0\leq p<\infty$},
 \label{40'} \\
 e^{itA} C_3 e^{-itA}&\leq& ke^{k'|t|} N^{1/2},
\label{40}
\end{eqnarray}
for some $0\leq k,k'<\infty$, and all $t\in\r$. Let $\psi$ be an eigenvector
of $L$. Then
there is a one-parameter family $\{\psi_\alpha\}\subset\dom(L)\cap \dom(C_1)$,
s.t. $\psi_\alpha\rightarrow\psi$, $\alpha\rightarrow 0$, and 
\begin{equation}
\lim_{\alpha\rightarrow 0}\scalprod{\psi_\alpha}{C_1\psi_\alpha}=0.
\label{42'}
\end{equation}
\end{theorem}

{\it Remarks.\ }\ 1) A sufficient condition for \fer{40'} to hold (with $k'=0$)
is that $N$
and $e^{itA}$ commute, for all $t\in\r$, in the strong sense on $\dom$, and $
C_1\leq kN^p$. This condition will always be 
satisfied in our applications. A similar remark applies to \fer{40}.\\
2) In a heuristic way, we understand $C_1$ as the commutator
$i[L,A]=i(LA-AL)$, and \fer{42'} as $\scalprod{\psi}{i[L,A]\psi}=0$, which is a
standard way of stating the virial theorem, see e.g. [ABG], and [GG] for a
comparison (and correction) of virial theorems encountered in the literature.\\
\indent
The result of the Virial Theorem is still valid if we add to the operator $A$ a suitably small perturbation $A_0$:\\

\begin{theorem}{\bf ($2^{\rm nd}$ virial theorem) }
\label{virialthm'}
\stepcounter{proposition}
 Suppose that we are in the situation of Theorem \ref{virialthm} and that
  $A_0$ is a bounded operator on $\h$ s.t. $\ran 
  A_0\subset\dom(L)\cap\ran P(N\leq n_0)$, for some $n_0<\infty$. Then
  $i[L,A_0]=i(LA_0-A_0L)$ is well defined in the strong sense on $\dom(L)$,
  and we have, for the same family of approximating eigenvectors as in Theorem
  \ref{virialthm}: 
\begin{equation}
\lim_{\alpha\rightarrow 0}\scalprod{\psi_{\alpha}}{(C_1+i[L,A_0])\psi_{\alpha}}=0.
\label{43}
\end{equation}
\end{theorem}

In conjunction with a positive commutator estimate, the virial theorem implies
a certain regularity of eigenfunctions. 

\begin{theorem}{\bf (Regularity of eigenfunctions)}
\label{roethm}
\stepcounter{proposition}
Suppose $C$ is a symmetric operator on a domain $\dom(C)$ s.t., in the sense of
quadratic  
forms on $\dom(C)$, we have that $C\geq {\cal P}-B$, where ${\cal P}\geq 0$ is a
selfadjoint operator, and $B$ is a bounded
(everywhere defined) operator. Let $\psi_\alpha$ be a family of vectors in
$\dom(C)$, with  
$\psi_\alpha\rightarrow \psi$, as $\alpha\rightarrow 0$, and s.t. 
\begin{equation}
\lim_{\alpha\rightarrow 0}\scalprod{\psi_\alpha}{C\psi_\alpha}=0.
\label{wtl}
\end{equation}
Then $\scalprod{\psi}{B\psi}\geq 0$, $\psi\in\dom({\cal P}^{1/2})$, and 
\begin{equation}
\|{\cal P}^{1/2}\psi\|\leq \scalprod{\psi}{B\psi}^{1/2}.
\label{wtl*}
\end{equation}
\end{theorem}

{\it Remark.\ } Theorem \ref{roethm} can be viewed as a consequence of an
abstract Fatou Lemma, see [ABG], Proposition 2.1.1. We give a different, very
short proof of \fer{wtl*} at the end of Section 6.

\subsection{The positive commutator method}
\label{pcmethod}

This method gives a conceptually very easy proof of absence of point
spectrum. The subtlety of the method lies in the technical details, since one
deals with unbounded operators. \\
\indent
Suppose we are in the setting of the virial theorems described in Section
\ref{absvthm}, and that the operator 
$C_1$ (or $C_1+i[L,A_0]$) is {\it strictly positive}, i.e.
\begin{equation}
C_1\geq\gamma,
\label{pc}
\end{equation}
for some $\gamma>0$. Inequality \fer{pc} and the virial theorem immediately
show that $L$ cannot have any eigenvalues. Indeed, assuming
$\psi$ is an eigenfunction of $L$, we reach the contradiction 
\begin{equation*}
0=\lim_{\alpha\rightarrow
  0}\scalprod{\psi_\alpha}{C_1\psi_\alpha}\geq \gamma \lim_{\alpha\rightarrow
  0}  \scalprod{\psi_\alpha}{\psi_\alpha}=\gamma\|\psi\|^2>0.
\end{equation*}
Although the {\it global PC estimate} \fer{pc} holds in our situation, often
one manages to prove merely a {\it localized} version. Suppose $g\in
C^\infty(J)$ is a smooth function with support in an interval $J\subseteq\r$,
 $g\upharpoonright J_1=1$, for some $J_1\subset J$, s.t. $g(L)$ leaves the
 form domain of $C_1$ invariant. The same reasoning as above shows that
 if
\begin{equation*}
g(L) C_1 g(L) \geq\gamma g^2(L),
\end{equation*}
for some $\gamma>0$, then $L$ has no eigenvalues in the interval $J_1$. The
use of PC estimates for spectral analysis of Schr\"odinger operators has
originated with Mourre [Mou], and had recent applications in [Ski], [BFSS],
[DJ], [Mer].

\section{Proof of Theorem \ref{mainthm}}
\label{mainthmproofsect}

\subsection{Strategy of the proof}
\label{ideaofproof}
As in [JP], [Mer], the starting point in 
the construction of a positive commutator is the adjoint operator
$A_f=\d\Gamma(i\partial_u)$, the second quantized generator of translation in
the radial variable of the glued Fock space $\cal F$, see \fer{calf}. We
formally have  
\begin{equation*}
i[L_0,A_f]=\d\Gamma(\bbbone_f)=N\geq 0.
\end{equation*}
 The kernel of this form is the
{\it infinite dimensional} space $\h_p\otimes\h_p\otimes\ran
P_\Omega$. Following [Mer], one is led to try to add a suitable operator 
$A_0$ to $A_f$, where $A_0$ 
depends on the interaction $\lambda I$, and is designed in such a way that 
$i[L_0+\lambda I, 
A_f+A_0]$ is strictly positive (has trivial kernel). This method is
applicable if 
the (imaginary part) of the so-called level shift operator is strictly
positive, or equivalently, if \fer{FGRC} is satisfied, but where the
finite-dimensional projection $p_0$ is replaced by the {\it
  infinite-dimensional} projection 
$\bbbone_p$. Such a positivity condition does not hold for reasonable
operators $G_\alpha$ and functions $g_\alpha$.\\
 \indent
In order to be able to carry out our program, we add to $A_f$ a term
$A_p\otimes\bbbone_p-\bbbone_p\otimes A_p$ that reduces the kernel of the commutator. A prime candidate for $A_p$ would
be the operator $i\partial_e$ acting on $\h_p$ (we write simply $i\partial_e$
instead of $0\oplus i\partial_e$, c.f. \fer{a16}), since then
\begin{equation*}
i[L_0,A_p\otimes\bbbone_p-\bbbone_p\otimes A_p+A_f]= P_+(H_p)\otimes\bbbone_p
+\bbbone_p\otimes P_+(H_p) +N,
\end{equation*}
where $P_+(H_p)=\dirint_{\r_+}de$ is the projection onto $L^2(\r_+,de;\h)$. The
above form has now a one-dimensional kernel, $\ran p_0\otimes p_0\otimes
P_\Omega$. By adding a suitable operator $A_0$, as described above, one can
obtain a 
lower bound on the commutator (and in particular, reduce
its kernel to $\{0\}$), provided \fer{FGRC} is satisfied.\\
\indent
However, the operator $A_p$ chosen above has the inconvenience of not being 
selfadjoint, while our virial theorems require
selfadjointness. We introduce a family of selfajoint
operators $A_p^a$, $a>0$, that approximate $i\partial_e$ in
a certain sense ($a\rightarrow 0$). The idea of approximating
a non-selfadjoint $A$ by a selfadjoint sequence was also used in
[Ski]. We now define $A^a$ and then explain, in the
remainder of this subsection, how to prove Theorem \ref{mainthm}.\\

We define $A_p^a$ as the generator of a unitary group on $L^2(\r_+,de;\h)$,
which is induced by a flow on $\r_+$. For the proof of the following
proposition, and more information on unitary groups induced by flows, we refer
to Section \ref{flowsection}.

\begin{proposition}
\label{xiprop}
\stepcounter{theorem}
Let $\xi:\r_+\rightarrow\r_+$ be a bounded, smooth vector field,
s.t. $\xi(0)=0$, $\xi(e)\rightarrow 1$, as
$e\rightarrow \infty$, and $\|(1+e)\xi'\|_\infty<\infty$. Then $\xi$ generates a global
flow, and this flow    
induces a continuous unitary group on $L^2(\r_+,de;\hh)$. The  generator $A_p$
of this group 
is essentially selfadjoint on $C_0^\infty$, and it acts on $C_0^\infty$ as 
\begin{equation}
A_p=i\left(\frac{1}{2}\xi'(e) +\xi(e)\partial_e\right), 
\label{a33}
\end{equation}
where  $\xi'(e)$ and $\xi(e)$ are multiplication operators. 
Given $a>0$, $\xi_a(e)=\xi(e/a)$ is a vector field on $\r_+$, and
$\lim_{a\rightarrow 0}\xi_a=1$, pointwise (except at zero). The generator
$A_p^a$ of the unitary group induced by $\xi_a$ is given on its core,
$C_0^\infty$, by 
\begin{equation}
A_p^a=i \left(\frac{1}{2}\frac{1}{a}\xi'(e/a) 
  +\xi(e/a)\partial_e\right).
\label{a36}
\end{equation}
\end{proposition}
 
We  define the selfadjoint operator
\begin{equation}
A^a=A_p^a\otimes\bbbone_p\otimes \bbbone_f -\bbbone_p\otimes
A^a_p\otimes\bbbone_f +A_f,
\label{Aa}
\end{equation}
and calculate the commutator $C_1^a$ of $iL$ with $A^a$ (in the sense given in
\fer{45}, see also Subsection \ref{concsetting}):
\begin{equation}
C_1^a=\dirint_{\r_+}\xi_a(e)de\otimes\bbbone_p+\bbbone_p\otimes
\dirint_{\r_+}\xi_a(e)de + N +\lambda I_1^a,
\label{a37}
\end{equation}
where $I_1^a$ is $N^{1/2}$-bounded. In Section \ref{masection}, we show
that $C^a_1+i[L,A_0]\geq M_a$, 
where $M_a$ is a bounded operator. We will see that 
$\mbox{s-}\lim_{a\rightarrow 0_+} M_a=M$ (see Proposition \ref{mtoma}), where $M$ is a
bounded, strictly positive  
operator (see Proposition \ref{positivemlemma}). Since $M_a$, $M$ are bounded,
we obtain from the virial theorem 
\begin{equation}
0=\lim_{\alpha\rightarrow
  0}\scalprod{\psi_\alpha}{(C_1^a+i[L,A_0])\psi_\alpha}\geq \scalprod{\psi}{(M_a-M)\psi}+\scalprod{\psi}{M\psi},
\label{a38}
\end{equation}
for any eigenfunction $\psi$ of $L$. Taking
$a\rightarrow 0_+$ and 
using strict positivity of $M$ (for small, but nonzero $\lambda$, see
Proposition 
\ref{positivemlemma}), gives a 
contradiction, and this will prove Theorem \ref{mainthm}.

\subsection{Concrete setting for the virial theorems}

\label{concsetting}
The Hilbert space is the GNS representation space \fer{7}, and we set
\begin{equation}
\dom=C_0^\infty\otimes C_0^\infty\otimes\dom_f,
\label{a39}
\end{equation}
where 
\begin{equation*}
\dom_f={\cal F}\left( C_0^\infty(\r\times S^2)\right)\cap {\cal F}_0,
\end{equation*}
and ${\cal F}_0$ denotes the finite-particle subspace of Fock space. The
operator 
$\Lambda$ is given by 
\begin{eqnarray}
\Lambda&=&\Lambda_p\otimes\bbbone_p+\bbbone_p\otimes\Lambda_p+\Lambda_f,\label{a40}\\ 
\Lambda_p&=&\dirint_{\r_+} e\ de+\bbbone_p=H_pP_+(H_p)+\bbbone_p,\label{a41}\\
\Lambda_f&=& \d\Gamma(u^2+1)+\bbbone_f.\label{a42}
\end{eqnarray}
In \fer{a41}, we have introduced $P_+(H_p)$, the projection onto the
spectral interval $\r_+$ of $H_p$. It is clear that $\Lambda$ is essentially
selfadjoint on $\dom$, and $\Lambda\geq\bbbone$.  The
operator $L$ is the interacting Liouvillian \fer{L}, and 
\begin{equation}
N=\d\Gamma(\bbbone)
\label{en}
\end{equation}
is the particle number operator in ${\cal F}\equiv {\cal F}(L^2(\r\times
S^2))$. Clearly, $X=L,N$ are 
symmetric operators on $\dom$, and the symmetric operator $D$ on $\dom$ (see
\fer{44}) is given by  
\begin{eqnarray}
D&=&\frac{i\lambda}{\sqrt{2}}\sum_\alpha\big\{G_{\alpha}\otimes\bbbone_p
    \otimes\left(-a^*(\tau_\beta(g_\alpha))+  
    a(\tau_\beta(g_\alpha))\right)\nonumber\\
&&-\bbbone_p\otimes\cc_p G_{\alpha}\cc_p \otimes\left(
    -a^*(e^{-\beta u/2}\tau_\beta(g_\alpha))+a(e^{-\beta
    u/2}\tau_\beta(g_\alpha))\right)\big\}.
\label{operatorD}
\end{eqnarray}
The operator $A$ is given by $A^a$ defined in \fer{Aa}. Notice that $A_p^a$
leaves $C_0^\infty$ 
invariant, $A_f$ leaves $\dom_f$ invariant, so $A^a$ maps $\dom$ into
$\dom(L)$. Furthermore, it is easy to see that $L$ maps $\dom$ into $\dom(A^a)$, hence
the commutator of $L$ with $A^a$ is well defined in the strong sense on
$\dom$. The same is true for the multiple commutators of $L$ with $A^a$.
 Setting $\xi'_a(e)=\xi'(e/a)$,
$\xi''_a(e)=\xi''(e/a)$, we obtain 
\begin{eqnarray}
C_1^a&=&\dirint_{\r_+}\xi_a(e)de\otimes\bbbone_p
+\bbbone_p\otimes\dirint_{\r_+}\xi_a(e)de +N+\lambda I_1^a,
\label{a43}\\
C_2^a&=&\frac{1}{a}\dirint_{\r_+}\xi'_a(e)\xi_a(e)de\otimes\bbbone_p
-\bbbone_p\otimes \frac{1}{a}\dirint_{\r_+}\xi'_a(e)\xi_a(e)de+\lambda I_2^a\label{a44}\\ 
C_3^a&=&\frac{1}{a^2}\dirint_{\r_+}\left(\xi_a''(e)\xi_a(e)^2
  +\xi'_a(e)^2\xi_a(e)\right)de\otimes\bbbone_p\nonumber\\
&&+\bbbone_p\otimes \frac{1}{a^2}\dirint_{\r_+}\left(\xi''_a(e)\xi_a(e)^2
  +\xi'_a(e)^2\xi_a(e)\right)de +\lambda I_3^a
\label{a45},
\end{eqnarray}
where 
\begin{eqnarray}
I^a_n&=&i^n\sum_{j=0}^n {n \choose k}\sum_\alpha\left\{ {\rm
    ad}_{A_p^a}^{(j)}(G_\alpha)\otimes\bbbone_p\otimes {\rm
    ad}_{A_f}^{(n-j)}\left(
    \varphi(\tau_\beta(g_\alpha))\right)\right.\nonumber\\
&&+\left.(-1)^j \bbbone_p\otimes{\rm
    ad}_{A_p^a}^{(j)}(\cc_p G_\alpha\cc_p )\otimes 
    {\rm ad}_{A_f}^{(n-j)}\left(\varphi(e^{\beta
    u/2}\tau_\beta(g_\alpha))\right)\right\},
\label{a46}
\end{eqnarray}
for $n=1,2,3$. \\

We define the bounded selfadjoint operator
$A_0$ on $\h$ by 
\begin{equation}
A_0=i\theta\lambda (\Pi
  I\repsilon^2\Pibar -\Pibar\repsilon^2I\Pi),\label{17}
\end{equation}
with 
\begin{equation}
\repsilon^2=(L_0^2+\epsilon^2)^{-1}.\label{18}
\end{equation}
Here, $\theta$ and $\epsilon$ are positive  parameters, and $\Pi$ is the
projection onto the zero eigenspace of $L_0$:
\begin{eqnarray}
\Pi&=&P_0\otimes P_\Omega,\label{19}\\
P_0&=&p_0\otimes p_0,\label{20}\\
\Pibar&=&\bbbone-\Pi\label{21},
\end{eqnarray}
where $p_0$ is the projection in ${\cal B}(\h_p)$ projecting onto the eigenspace corresponding to the
eigenvalue $E$ of $H_p$, i.e. $p_0\psi=\psi(E)\in\cx$, and $P_\Omega$ is the projection in ${\cal B}({\cal
  F})$ projecting onto ${\mathbb C}\Omega$. We also introduce the notation
\begin{equation*}
\repsilonbar=\Pibar \repsilon.
\end{equation*}
\indent
Notice that the operator $A_0$ satisfies the conditions given in  Theorem
\ref{virialthm'} with $n_0=1$. Moreover, $[L,A_0]=LA_0-A_0L$ extends to a bounded operator on the entire Hilbert space, and 
\begin{equation}
\|[L,A_0]\|\leq k\left(\frac{\theta\lambda}{\epsilon}+\frac{\theta\lambda^2}{\epsilon^2}\right).
\label{a}
\end{equation}
\indent
This choice for the operator $A_0$ was initially introduced in [BFSS] for the
spectral analysis of Pauli-Fierz Hamiltonians (zero temperature systems), and
was adopted in [Mer] to show return to equilibrium (positive temperature
systems). The key feature of $A_0$ is that $i\Pi[L,A_0]\Pi=2\theta\lambda^2
\Pi I\repsilonbar^2 I\Pi$ is a {\it non-negative} operator. Assuming the Fermi
Golden Rule Condition \fer{FGRC}, it is a {\it
  strictly positive operator}, as shows
\begin{proposition}
\label{fgrprop}
\stepcounter{theorem}
Assume Condition (A3). For $0<\epsilon<\epsilon_0$, we have 
\begin{equation}
\Pi I\repsilonbar^2 I\Pi\geq \frac{\gamma}{\epsilon}\Pi.
\label{fgrc1}
\end{equation}
\end{proposition}
The proof is given in Section \ref{propproofsection}.\\
\indent
 We are now ready to
verify that the virial theorems are applicable. 

\begin{proposition}
\label{alemma}
\stepcounter{theorem}
The unitary group $e^{itA^a}$ leaves $\dom(\Lambda)$ invariant ($a>0,
t\in\r$), and, for $\psi\in\dom(\Lambda)$,
\begin{equation}
\|\Lambda e^{itA^a}\psi\|\leq k e^{k'|t|/a}\|\Lambda\psi\|,
\label{a47}
\end{equation}
where $k,k'<\infty$ are independent of $a$.
\end{proposition}
The proof is given in Section \ref{propproofsection}.\\
\indent
Next, we verify the GJN conditions, and the bounds \fer{41}, \fer{40},
\fer{40'}. The following result is useful.

\begin{proposition}
\label{boundedcomms}
\stepcounter{theorem}
Under conditions \fer{regul1}, \fer{regul2},  the multiple commutators of
$G_\alpha$ with $A^a_p$ are well defined 
in the strong sense on $C_0^\infty$, and, for any $\psi\in
C_0^\infty$, we have that
\begin{equation}
\|{\rm ad}_{A^a_p}^{(n)}(G_\alpha)\psi\|\leq k\|\psi\|,
\label{a48}
\end{equation}
for $n=1,2,3$, and uniformly in $a>0$.
\end{proposition}
The proofs of this and the next proposition are given in Section
\ref{propproofsection}.  

\begin{proposition}
\label{vtapply}
\stepcounter{theorem}
The virial theorems, Theorem \ref{virialthm} and Theorem \ref{virialthm'},
apply in the concrete
situation described above, with the following identifications: the
domain $\dom$ of section \ref{absvthm} is 
given in \fer{a39}, the operators $L, N, D, \Lambda, A_0$ appearing in
Theorems \ref{virialthm}, \ref{virialthm'} are chosen in
\fer{L}, \fer{en}, \fer{operatorD}, \fer{a40}, \fer{17}, and the operator $A$
is given by $A^a$ in \fer{Aa}.
\end{proposition}

\subsection{A lower bound on $C_1^a+i[L,A_0]$ uniform in $a$}
\label{masection}
In order to estimate $C_1^a+i[L,A_0]$ from below, we start with the
following observation: in the sense of forms on $\dom$,
\begin{equation}
\pm\lambda I_1^a\leq \frac{1}{10} N\Pbar_\Omega +k\lambda^2,
\label{a49}
\end{equation}
for some $k$ independent of $a>0$. This estimate follows in a standard way
from the explicit expression for $I_1^a$, equation \fer{a46}, and the bound in 
\fer{a48}. We 
conclude from \fer{a49}, \fer{a43} that 
\begin{equation}
C_1^a+i[L,A_0]\geq M_a,
\label{a60}
\end{equation}
where 
\begin{equation}
M_a=\dirint_{\r_+}\xi_a(e)de\otimes\bbbone_p
+\bbbone_p\otimes\dirint_{\r_+}\xi_a(e)de
+\frac{9}{10}\Pbar_\Omega-k\lambda^2+i[L,A_0].
\label{a61}
\end{equation}
The constant $k$ on the r.h.s. is independent of
$a$. Recalling that $\xi_a\rightarrow 1$ a.e., we are led to define the
bounded limiting operator
\begin{equation}
M=P_+(H_p)\otimes\bbbone_p+\bbbone_p\otimes P_+(H_p)+\frac{9}{10}\Pbar_\Omega
-k\lambda^2 +i[L,A_0],
\label{a62}
\end{equation}
where $k$ is the same constant as in \fer{a61}. Using Dominated Convergence,
 one   
 readily verifies that $\dirint_{\r_+}\xi_a(e)de\rightarrow
 P_+(H_p)$, in the strong sense on $\h_p$. 
\begin{proposition}
\label{mtoma}
\stepcounter{theorem}
$\lim_{a\rightarrow 0_+}M_a=M$, strongly on $\h$.
\end{proposition}
Our next task is to show that $M$ is strictly positive.

\subsection{The Feshbach method and strict positivity of $M$}
\label{feshsection}
Recall that $\Pi=P_0\otimes P_\Omega$ is the rank-one projection onto the zero
eigenspace of $L_0$, see \fer{19}. We apply the Feshbach method to analyze the
operator $M$, with the decomposition 
\begin{equation*}
\h=\ran\Pi\oplus\ran\Pibar.
\end{equation*}
First, we note that 
\begin{eqnarray}
\Pibar M\Pibar&\geq&\Pbar_0\otimes P_\Omega\left( P_+(H_p)\otimes\bbbone_p
  +\bbbone_p \otimes P_+(H_p) -k\lambda^2\right)\nonumber\\
&& +(9/10-k\lambda^2)\Pbar_\Omega+i\Pibar[L,A_0]\Pibar.
\label{a64}
\end{eqnarray}
Recalling the definitions of $P_0$ and $A_0$, \fer{20} and \fer{17}, one
easily sees that 
\begin{eqnarray*}
\Pbar_0(P_+(H_p)\otimes\bbbone_p+\bbbone_p\otimes P_+(H_p))&\geq&
\Pbar_0,\\
 i\Pibar[L,A_0]\Pibar&=&-\theta\lambda^2(\Pibar I\Pi I\repsilonbar^2
+\repsilonbar^2 I\Pi I\Pibar),
\end{eqnarray*}
in particular, $\|i\Pibar[L,A_0]\Pibar\|\leq k\theta\lambda^2/\epsilon^2$. Together
with \fer{a64}, this shows that there is a constant $\lambda_1>0$ (independent
of $\lambda,\theta,\epsilon$ and of $\beta\geq \beta_0$, for any $\beta_0>0$
fixed), s.t.  
\begin{equation}
\Mbar:=\Pibar M\Pibar\upharpoonright\ran\Pibar > \frac{1}{2}\Pibar,
\label{a65}
\end{equation}
provided
\begin{equation}
|\lambda|, \frac{\theta\lambda^2}{\epsilon^2} < \lambda_1.
\label{a66}
\end{equation}
It follows from equation \fer{a65} that the resolvent set of $\Mbar$,
$\rho(\Mbar)$, contains the interval $(-\infty,1/2)$, and for $m<1/2$:
\begin{equation}
\|(\Mbar-m\Pibar)^{-1}\|<(1/2-m)^{-1}.
\label{a67}
\end{equation}
For $m\in\rho(\Mbar)$, we define the {\it Feshbach map $F_{\Pi,m}$} applied to
$M$ by
\begin{equation}
F_{\Pi,m}(M)=\Pi\left( M-M\Pibar(\Mbar-m\Pibar)^{-1}\Pibar M\right)\Pi.
\label{a68}
\end{equation}
The operator $F_{\Pi,m}(M)$ acts on the space $\ran\Pi$. In our
specific case, $\ran\Pi\cong\cx$, hence $F_{\Pi,m}(M)$ is a number. (If
$\ran\Pi$ had dimension $n$, then $F_{\Pi,m}(M)$ would be represented by an
$n\times n$ matrix.) The following crucial property is called the {\it
  isospectrality of the Feshbach map} (see e.g. [BFS], [DJ]):
\begin{equation}
m\in\rho(\Mbar)\cap\sigma(M)\Longleftrightarrow m\in\rho(\Mbar)\cap
\sigma(F_{\Pi,m}(M)),
\label{a69}
\end{equation}
where $\sigma(\cdot)$ denotes the spectrum. Hence by examining the spectrum of
the operator $F_{\Pi,m}(M)$, one obtains information about the spectrum of
$M$. The 
idea is, of course, that it is easier to examine the former operator, since it
acts on a smaller space.
\begin{proposition}
\label{feshlemma}
\stepcounter{theorem}
Assume condition (A3) and let $0<\epsilon<\epsilon_0$. Then 
\begin{equation}
F_{\Pi,m}(M)\geq
2\frac{\theta\lambda^2}{\epsilon}\gamma\left(1-k\theta\left(1+\frac{|\lambda|}{\epsilon}\right)^2-k\frac{\epsilon}{\gamma\theta}\right) \Pi, 
\label{a70}
\end{equation}
uniformly in $m<1/4$. 
\end{proposition}

{\it Proof.\ } Recall the structure of $F_{\Pi,m}(M)$, given in \fer{a68}. We
show that $-\Pi M\Pibar(\Mbar-m\Pibar)^{-1}\Pibar M\Pi$ is 
small, as compared to $\Pi M\Pi$, and that the latter is strictly positive. Estimate \fer{a67} gives
\begin{equation}
-\Pi M\Pibar (\Mbar-m\Pibar)^{-1}\Pibar M\Pi\geq -4\Pi M\Pibar M\Pi,
\label{a71}
\end{equation}
for $m<1/4$. An easy calculation shows that 
\begin{equation*}
\Pibar M\Pi=\Pibar i[L,A_0]\Pi =\theta\lambda\Pibar L\repsilonbar^2 I\Pi
=\theta\lambda\Pibar \left( L_0\repsilonbar^2 I+\lambda I\repsilonbar^2
  I\right) \Pi,
\end{equation*}
and using that $\|L_0\repsilon\|\leq 1$, $\|\repsilon\|\leq 1/\epsilon$, we
obtain the bound
\begin{equation}
\|\Pibar M\Pi\psi\|\leq
\left(\theta|\lambda|+k\frac{\theta\lambda^2}{\epsilon}\right)\|\repsilonbar
I\Pi\psi\|,
\label{a72}
\end{equation}
for any $\psi\in\h$, 
where we have used that $\ran \repsilonbar^2I\Pi\subset\ran P(N\leq 1)$, and
$\|IP(N\leq 1)\|\leq k$. Combining \fer{a72} with \fer{a71} yields
\begin{equation*}
-\Pi M\Pibar (\Mbar-m\Pibar)^{-1}\Pibar M\Pi\geq -k\theta^2\lambda^2
 (1+|\lambda|/\epsilon)^2 \Pi I\repsilonbar^2I\Pi.
\end{equation*}
Furthermore, we have that 
\begin{equation*}
\Pi M\Pi =\Pi i[L,A_0]\Pi-k\lambda^2 \Pi=2\theta\lambda^2 \Pi I\repsilonbar^2
I\Pi-k\lambda^2 \Pi.
\end{equation*}
These observations and  the definition of the Feshbach map, \fer{a68}, show
that  
\begin{equation*}
F_{\Pi,m}(M)\geq 2\theta\lambda^2(1-k\theta(1+|\lambda|/\epsilon)^2)\Pi
I\repsilonbar^2 I\Pi-k\lambda^2\Pi,
\end{equation*}
which, by Proposition \ref{fgrprop}, yields \fer{a70}.\hfill $\blacksquare$
\\ 

Estimate \fer{a70} tells us that there is a $\lambda_2>0$ s.t. 
\begin{equation}
F_{\Pi,m}(M)\geq \frac{\theta\lambda^2}{\epsilon}\gamma\Pi,
\label{a73}
\end{equation}
provided conditions \fer{a66} hold, and 
\begin{equation}
\theta\left(1+\frac{|\lambda|}{\epsilon}\right)^2
+\frac{\epsilon}{\gamma\theta} <\lambda_2,\ \ 0<\epsilon<\epsilon_0. 
\label{a74}
\end{equation}
Notice that all these estimates are independent of $m<1/4$. Using the
isospectrality property of the Feshbach map, \fer{a69}, we conclude that if
the bounds \fer{a66} and \fer{a74} are imposed on the parameters, and if
$m<1/4$ 
and $m\in\sigma(M)$, then $m>\frac{\theta\lambda^2}{\epsilon}\gamma$.
Consequently, 
\begin{equation*}
M\geq\min\left\{\frac{1}{4},\frac{\theta\lambda^2}{\epsilon}\gamma\right\}=\frac{\theta\lambda^2}{\epsilon}\gamma. 
\end{equation*}
Fix a $\theta<\lambda_2/4$ and an $\epsilon<\min\{\epsilon_0,
 \gamma\theta\lambda_2\}$. Then, defining  
\begin{equation*}
\lambda_0=\min\left\{\lambda_1,\frac{\epsilon\sqrt{\lambda_1}}{\sqrt{\theta}},
 \epsilon \right\},
\end{equation*}
\fer{a66} and \fer{a74} are satisfied for $|\lambda|<\lambda_0$. 
\begin{proposition}
\label{positivemlemma}
\stepcounter{theorem}
There is a choice of the parameters $\theta$ and $\epsilon$, and of
$\lambda_0>0$ (depending on $\theta,\epsilon,\beta$) s.t. if $|\lambda|<\lambda_0$
then
\begin{equation}
M> \frac{\theta\lambda^2}{\epsilon}\gamma.
\label{a75}
\end{equation}
We have $\lambda_0\leq k\gamma$, for some $k$ independent of
$\beta\geq\beta_0$ (for any $\beta_0>0$ fixed), i.e., $\lambda_0\sim e^{\beta
  E}$ is
exponentially small in $\beta$, as $\beta\rightarrow\infty$ (see remark 1)
after \fer{FGRC}).  
\end{proposition}

Proposition \ref{positivemlemma} completes the  proof of Theorem
\ref{mainthm}, according to the argument given in \fer{a38}.

\section{Some functional analysis}
\label{sfa}

The following two theorems are useful in our analysis. Their
proofs can be found in [Fr\"o]. 

\begin{theorem}{\bf (invariance of domain, [Fr\"o]) }
\label{IDthm}
\stepcounter{proposition}
 Suppose $(X,Y,\dom)$ satisfies
    the GJN Condition, \fer{nc1}, \fer{nc2}. Then the unitary group,
    $e^{itX}$, generated by the selfadjoint operator 
    $X$ leaves $\dom(Y)$ invariant, and 
\begin{equation}
\|Ye^{itX}\psi\|\leq e^{k|t|}\|Y\psi\|,
\label{80}
\end{equation}
for some $k\geq 0$, and all $\psi\in\dom(Y)$.
\end{theorem}

\begin{theorem}{\bf (commutator expansion, [Fr\"o]) }
\label{CEthm}
\stepcounter{proposition}
 Suppose $\dom$ is a core
  for the selfadjoint operator $Y\geq\bbbone$. Let $X,Z, {\rm ad}_X^{(n)}(Z)$
  be symmetric operators on $\dom$, where 
\begin{eqnarray*}
{\rm ad}_X^{(0)}(Z)&=&Z,\\
\scalprod{\psi}{{\rm ad}_X^{(n)}(Z)\psi}&=&i\left\{\scalprod{{\rm
      ad}_X^{(n-1)}(Z)\psi}{X\psi} -\scalprod{X\psi}{{\rm
      ad}_X^{(n-1)}(Z)\psi}\right\},
\end{eqnarray*}
for all $\psi\in\dom$, $n=1,\ldots,M$. We suppose that the triples
$({\rm ad}_X^{(n)}(Z),Y,\dom)$, $n=0,1,\ldots,M$, satisfy the
GJN Condition \fer{nc1}, \fer{nc2}, and that $X$ is 
selfadjoint, with $\dom\subset\dom(X)$, $e^{itX}$ leaves $\dom(Y)$ invariant,
and \fer{80} holds. Then
\begin{eqnarray}
e^{itX}Ze^{-itX}&=& Z-\sum_{n=1}^{M-1}\frac{t^n}{n!}
{\rm ad}_X^{(n)}(Z)\nonumber\\
&&-\int_0^tdt_1\cdots\int_0^{t_{M-1}} dt_M e^{it_MX}{\rm
  ad}_X^{(M)}(Z)e^{-it_MX},
\label{cmexp}
\end{eqnarray}
as operators on $\dom(Y)$. 
\end{theorem}

{\it Remark.\ } This theorem is proved in [Fr\"o], under the assumption that
$(X,Y,\dom)$ satisfies \fer{nc1}, \fer{nc2}. However, [Fr\"o]'s proof only
requires the properties of the group $e^{itX}$ indicated in our
Theorem \ref{CEthm}.\\
\indent
An easy, but useful result follows from \fer{80}.

\begin{proposition}
\label{easybutusefullemma}
\stepcounter{theorem}
Suppose that the unitary group $e^{itX}$ leaves $\dom(Y)$ invariant, for some
operator $Y$, and that estimate \fer{80} holds. For a 
function $\chi$ on $\r$ with Fourier transform $\what{\chi}\in L^1(\r)$, we
define 
$\chi(X)=\int_\r \what{\chi}(s) e^{isX}ds$. If $\what{\chi}$ has compact
support, then $\chi(X)$ leaves $\dom(Y)$ invariant, and, for $\psi\in\dom(Y)$,
\begin{equation} 
\|Y\chi(X)\psi\|\leq e^{k R}\|\what{\chi}\|_{L^1(\r)}\ \|Y\psi\|,
\label{easybutusefullabel}
\end{equation}
for any $R$ s.t. $\supp\what{\chi}\subset [-R,R]$.
\end{proposition}

The proof is obvious. Proposition \ref{invlemma} states a similar
result, but for a function whose Fourier transform is not necessarily of
compact support.

\begin{proposition}
\label{invlemma}
\stepcounter{theorem}
Suppose $(X,Y,\dom)$ satisfies the GJN Condition, and so
  do the triples $({\rm ad}_X^{(n)}(Y),Y,\dom)$, for $n=1,\ldots,M$, and for
  some $M\geq 
  1$. Moreover, assume that, in the sense of Kato on $\dom(Y)$, 
  $\pm{\rm ad}_X^{(M)}(Y)\leq k X$, for some $k\geq 0$. For $\chi\in
  C_0^\infty(\r)$, a smooth function with compact support,  define
  $\chi(X)=\int\what{\chi}(s)e^{isX}$, where $\what{\chi}$ is the Fourier
  transform of $\chi$. Then $\chi(X)$ leaves $\dom(Y)$ invariant.
\end{proposition}

{\it Proof.\ }
For $R>0$, set $\chi_R(X)=\int_{-R}^R\what{\chi}(s)e^{isX}$, then
$\chi_R(X)\rightarrow \chi(X)$ in operator norm, as $R\rightarrow
\infty$. From the invariance of domain theorem, we see that $\chi_R(X)$ leaves
$\dom(Y)$ invariant. Let $\psi\in\dom(Y)$, then using the commutator expansion
theorem above, we have
\begin{eqnarray}
\lefteqn{
Y\chi_R(X)\psi}\nonumber\\
&=&\chi_R(X)Y\psi +\int_{-R}^R\what{\chi}(s)
    e^{isX}\left(e^{-isX} Y e^{isX}-Y \right)\psi\nonumber\\
&=&\chi_R(X)Y\psi-\int_{-R}^R\what{\chi}(s)e^{isX}\left(\sum_{n=1}^{M-1}\frac{(-s)^n}{n!}{\rm
    ad}_X^{(n)}(Y)\right.\nonumber\\
&&\left.+(-1)^M\int_0^sds_1\cdots\int_0^{s_{M-1}}ds_Me^{-is_MX}{\rm
    ad}_X^{(M)}(Y) e^{is_MX}\right)\psi.
\label{81}
\end{eqnarray}
The integrand of the $s$-integration in \fer{81} is bounded in norm by 
\begin{equation*}
k(|s|^M+1)\left(\|Y\psi\|+\|X\psi\|\right)\leq k(|s|^M+1)\|Y\psi\|,
\end{equation*}
where we have used that $\|{\rm ad}_X^{(M)}(Y)e^{is_MX}\psi\|\leq
\|Xe^{is_MX}\psi\|\leq \|X\psi\|$. Since $\what{\chi}$ is of rapid decrease, it
can be integrated against any power of $|s|$, and we conclude that the
r.h.s. of \fer{81} has a limit as $R\rightarrow\infty$. Since $Y$ is a closed
operator, it follows that $\chi(X)\psi\in\dom(Y)$.\hfill $\blacksquare$

\begin{proposition}
\label{boundedcommlemma}
\stepcounter{theorem}
Let $\chi\in C_0^\infty(\r)$, $\chi=F^2\geq 0$. Suppose $(X,Y,\dom)$
satisfies the GJN condition. Suppose $F(X)$ leaves $\dom(Y)$
invariant. Let $Z$ be a symmetric operator on $\dom$ s.t., for some $M\geq 1$,
and $n=0,1,\ldots,M$, the triples $({\rm ad}_X^{(n)}(Z),Y,\dom)$ satisfy the
GJN condition. Moreover, we assume that the multiple commutators, for
$n=1,\ldots,M$, are 
relatively $X^{2p}$-bounded in the sense of Kato on $\dom$, for some $p\geq 0$. In other words, there
is some $k<\infty$, s.t. $\forall \psi\in \dom$,
\begin{equation*}
\|{\rm ad}_X^{(n)}(Z)\psi\|\leq k\left( \|\psi\|+\|X^{2p}\psi\|\right) ,\ \ \ \ n=1,\ldots,M.
\end{equation*}
Then the commutator $[\chi(X),Z]=\chi(X)Z-Z\chi(X)$ is well defined on $\dom$
and extends to a bounded operator.
\end{proposition}

{\it Proof.\ }
We write $F,\chi$ instead of $F(X), \chi(X)$. Since $F$ leaves $\dom(Y)$
invariant, we have that
\begin{equation*}
[\chi,Z]=F[F,Z]+[F,Z]F,
\end{equation*}
as operators on $\dom(Y)$. 
We expand the commutator
\begin{eqnarray}
[F,Z]&=&\int\what{F}(s) e^{isX}\left( Z-e^{-isX}Ze^{isX}\right)\nonumber\\
&=&\int\what{F}(s)e^{isX}\left\{\sum_{n=1}^{M-1}\frac{s^n}{n!}{\rm
    ad}_X^{(n)}(Z)\right.\nonumber\\
&&\ \ \ \left.+ \int_0^sds_1\cdots\int_0^{s_{M-1}}ds_M e^{-is_MX} {\rm
    ad}_X^{(M)}(Z) e^{is_MX}\right\}.
\label{juerg's}
\end{eqnarray}
Multiplying this equation from the right with $F$ (and noticing that $F$
commutes with 
$e^{is_MX}$), we see immediately that $[F,Z]F$ is bounded, and hence
$F[F,Z]=-([F,Z]F)^*$ is bounded, too.\hfill $\blacksquare$

\begin{proposition}
\label{reslemma}
\stepcounter{theorem}
Suppose $(X,Y,\dom)$ is a GJN triple. Then the
resolvent $(X-z)^{-1}$ leaves $\dom(Y)$ invariant, for all $z\in\{{\mathbb C} |
\ |\IM z|>k \}$, for some $k>0$.
\end{proposition}

{\it Proof.\ }Suppose $\IM z<0$ (the case $\IM z>0$ is dealt with
similarly). We write the resolvent as
\begin{equation*}
(X-z)^{-1}=i \int_0^\infty dt \ e^{i(X-z)t},
\end{equation*}
and it follows from Theorem \ref{IDthm} that for $\psi\in\dom(Y)$,
\begin{equation*}
\|Y(X-z)^{-1}\psi\|\leq \|Y\psi\|\int_0^\infty dt\ e^{(\IM z+k)t}<\infty,
\end{equation*}
provided $\IM z<-k$.\hfill $\blacksquare$

\section{Proof of the virial theorems and the regularity theorem}
\label{vtproof}

{\it Proof of Theorem \ref{virialthm}.\ }
We start by introducing some cutoff operators, and the regularized (cutoff,
approximate) eigenfunction.\\
\indent
Let $g_1 \in C_0^\infty( (-1,1) )$ be a real valued function,
s.t. $g_1(0)=1$, and set $g=g_1^2\in 
C_0^\infty( (-1,1) )$, $g(0)=1$. Pick a real valued function $f$ on $\r$ with
the properties that $f(0)=1$ and $\what{f}\in C_0^\infty(\r)$ (Fourier
transform). We set
\begin{equation*}
f_1(x)=\int_{-\infty}^x f^2(y)dy,
\end{equation*}
so that $f_1'(x)=f^2(x)$. Since $\what{f_1'\
  }(s)=is\what{f_1}(s)= (2\pi)^{-1/2}\what{f}*\what{f}(s)$, it follows that
  $\what{f_1}$ has compact support, and is smooth except at $s=0$, where it
  behaves like $s^{-1}$. We have
  $\what{f^{(n)}_1}=(is)^n\what{f_1}\in C_0^\infty$, for $n\geq 1$.
 Let
$\alpha, \nu>0$ be two parameters and define the cutoff-operators
\begin{eqnarray*}
g_{1,\nu}&=&g_1(\nu N)=\int_\r \what{g_1}(s) e^{is\nu N} ds,\\
g_\nu&=&g_{1,\nu}^2\\
f_\alpha&=& f(\alpha A)=\int_\r \what{f}(s) e^{is\alpha A}ds,\\
\end{eqnarray*}
For $\eta>0$, define
\begin{equation*}
f^\eta_{1,\alpha}=\frac{1}{\alpha}\int_{\r\backslash (-\eta,\eta)} ds
\what{f_1}(s) e^{is\alpha A}= (f^\eta_{1,\alpha})^*.
\end{equation*}
$f^\eta_{1,\alpha}$ leaves $\dom(\Lambda)$ invariant, and
$\|f^\eta_{1,\alpha}\|\leq k/\alpha$, where $k$ is a constant independent of
$\eta$; this can be seen by noticing that $\|f_1\|_\infty<\infty$. \\
\indent
Suppose that $\psi$ is an eigenfunction of $L$ with eigenvalue $e$:
$L\psi=e\psi$.
Since $\psi\in\dom(L)$, then
$\psi=(L+i)^{-1}\varphi$, for some $\varphi\in\h$. Let
$\{\varphi_n\}\subset\dom(\Lambda)$ be a sequence of vectors converging to
$\varphi$. Then
\begin{equation}
\psi_n:=(L+i)^{-1}\varphi_n\rightarrow\psi,\ \ \ n\rightarrow\infty,
\label{*}
\end{equation}
 and moreover, $\psi_n\in\dom(\Lambda)$. The latter follows because
the 
resolvent $(L+i)^{-1}$ leaves $\dom(\Lambda)$ invariant, see Proposition
\ref{reslemma}; without loss of generality, we assume that $k=1$. Moreover, by
Proposition \ref{easybutusefullemma}, we know that $f_\alpha$ leaves
$\dom(\Lambda)$ invariant (see also \fer{alambda}), and $g_\nu$ leaves
$\dom(\Lambda)$ invariant ($\Lambda$ commutes with $N$ in the strong sense on $\dom$). Hence, the
regularized eigenfunction 
\begin{equation*}
\psi_{\alpha,\nu,n}=f_\alpha g_\nu\psi_n
\end{equation*}
satisfies $\psi_{\alpha,\nu,n}\in\dom(\Lambda)$,
$\psi_{\alpha,\nu,n}\rightarrow\psi$, as $\alpha, \nu\rightarrow 0$,
$n\rightarrow \infty$. \\
\indent
Notice that in the definition of $\psi_n$, we introduced the resolvent of $L$,
so that we have $(L-e)\psi_n\rightarrow 0$, as $n\rightarrow\infty$, which we
write as 
\begin{equation}
(L-e)\psi_n=o(n).
\label{o1n}
\end{equation}
We  now prove the  estimate
\begin{equation}
\left|\av{if^\eta_{1,\alpha}(L-e)}_{g_\nu\psi_n}\right|\leq
k\frac{1}{\alpha}\left(\sqrt{\nu} +o(n)\right),
\label{a0}
\end{equation}
where $k$ is some constant independent of $\eta,\alpha,\nu, n$. This estimate
follows from the bound
\begin{equation}
\|(L-e)g_\nu\psi_n\|\leq k\left(\sqrt{\nu} +o(n)\right),
\label{a1}
\end{equation}
which is proven as follows. We have that
\begin{eqnarray}
(L-e)g_\nu\psi_n&=& g_\nu (L-e)\psi_n\label{a2}\\
&&+g_{1,\nu}[L,g_{1,\nu}]\psi_n\label{a3}\\
&&+[L,g_{1,\nu}]g_{1,\nu}\psi_n,\label{a4}
\end{eqnarray}
and the r.h.s. of \fer{a2} is $o(n)$, by \fer{o1n}.
 Let us show that both \fer{a3} and \fer{a4} are bounded above by
 $k\sqrt{\nu}$, uniformly in $n$. The commutator expansion of Theorem
 \ref{CEthm} (see also \fer{juerg's}) yields 
\begin{equation}
g_{1,\nu}[L,g_{1,\nu}]=\nu\int_\r ds\ \what{g_1}(s) e^{is\nu N}\int_0^s
ds_1e^{-is_1\nu N}g_{1,\nu}D e^{is_1\nu N}, 
\label{a5'}
\end{equation}
as operators on  $\dom(\Lambda)$, where $D$ is given in \fer{operatorD}. We
use that $g_{1,\nu}$ commutes with $e^{is\nu N}$. 
From \fer{41}, we see that for any $\phi\in\dom(\Lambda)$,
\begin{eqnarray*}
\lefteqn{\|g_{1,\nu}De^{is_1\nu N}\phi\|=\sup_{\varphi\in \dom, \varphi\neq
  0}\frac{\left|\scalprod{\varphi}{g_{1,\nu}De^{is_1\nu
  N}\phi}\right|}{\|\varphi\|}}\\
&\leq&  
\sup_{\varphi\in \dom, \varphi\neq
  0}\frac{\|Dg_{1,\nu}\varphi\|\ \|\phi\|}{\|\varphi\|}
\leq k \sup_{\varphi\in \dom, \varphi\neq
  0}\frac{\|N^{1/2}g_{1,\nu}\varphi\|}{\|\varphi\|}\ \|\phi\|
\leq k\frac{1}{\sqrt{\nu}}\|\phi\|,
\end{eqnarray*}
and consequently,
\begin{eqnarray}
\|g_{1,\nu}[L,g_{1,\nu}]\phi\|&\leq&\nu\int_\r ds |\what{g_1}(s)|\int_0^s ds_1\,
\|g_{1,\nu}De^{is_1\nu N}\phi\|\nonumber\\
&\leq & k\sqrt{\nu}\int_\r ds\, |s\what{g_1}(s)| \ \|\phi\|.
\label{a5}
\end{eqnarray}

Thus, the desired bound for \fer{a3} is proven, and the same bound is established for 
\fer{a4} by proceeding in a similar way. This proves \fer{a1}.\\
\indent
Next, since $f^\eta_{1,\alpha}$ leaves
$\dom(\Lambda)$ invariant, the commutator $[f^\eta_{1,\alpha},L]$ is 
defined in the strong sense on $\dom(\Lambda)$, and Theorem \ref{CEthm}
yields
\begin{eqnarray}
\lefteqn{[f^\eta_{1,\alpha},L]}\nonumber\\
&=&\int_{\r\backslash (-\eta,\eta)} ds \what{f_1}(s)e^{is\alpha A}\left(sC_1
  +\alpha\frac{s^2}{2}C_2\right)\nonumber\\
&&+\alpha^2\int_{\r\backslash (-\eta,\eta)} ds\what{f_1}(s) e^{is\alpha
  A}\int_0^sds_1\int_0^{s_1}ds_2\int_0^{s_2}ds_3 \ e^{-is_3\alpha
  A}C_3e^{is_3\alpha A}.\nonumber\\
\label{a6'}
\end{eqnarray}
For $n\geq 1$, we have
\begin{equation*}
f_1^{(n)}(\alpha A)=\int_\r ds (is)^n\what{f_1}(s) e^{is\alpha
  A}=\int_{\r\backslash (-\eta,\eta)} ds (is)^n\what{f_1}(s) e^{is\alpha
  A}-{\cal R}_{\eta, n},
\end{equation*}
where the remainder term 
\begin{equation*}
{\cal R}_{\eta,n}=-\int_{-\eta}^\eta ds\ (is)^n \what{f_1}(s) e^{is\alpha A}
\end{equation*}
satisfies ${\cal R}_{\eta,n}=({\cal R}_{\eta,n})^*$, and $\|{\cal
  R}_{\eta,n}\|\leq k_n\eta$, with a constant $k_n$ that does not depend on
  $\alpha,\eta$. We obtain from \fer{a6'}
\begin{eqnarray}
[f^\eta_{1,\alpha},L]&=&-i\left(f_1'(\alpha A)+{\cal
  R}_{\eta,1}\right)C_1-\frac{\alpha}{2}\left( f''_1(\alpha A)+{\cal
  R}_{\eta,2}\right)C_2\nonumber\\
&&+\alpha^2\int_{\r\backslash (-\eta,\eta)} ds\what{f_1}(s) e^{is\alpha
  A}\int_0^sds_1\int_0^{s_1}ds_2\int_0^{s_2}ds_3 \ e^{-is_3\alpha
  A}C_3e^{is_3\alpha A}.\nonumber\\
\label{a6}
\end{eqnarray}
Recalling that $f'_1(\alpha A)=f^2(\alpha A)=f_\alpha^2$, we write 
\begin{eqnarray}
-if_\alpha^2C_1&=& -if_\alpha C_1f_\alpha\nonumber\\
&&-if_\alpha\int_\r ds\what{f}(s) e^{is\alpha A}\left(\alpha s
  C_2+\alpha^2\int_0^sds_1\int_0^{s_1}ds_2 e^{-is_2\alpha A} C_3 e^{is_2\alpha
    A}\right)\nonumber\\
&=&-if_\alpha C_1f_\alpha -\alpha f_\alpha f'_\alpha C_2\nonumber\\
&&-i\alpha^2 f_\alpha \int_\r ds\what{f}(s) e^{is\alpha A}
 \int_0^sds_1\int_0^{s_1}ds_2 e^{-is_2\alpha A} C_3 e^{is_2\alpha
    A},
\label{a8}
\end{eqnarray}
where $f_\alpha'=f'(\alpha A)$. Remarking that 
 $f_\alpha f'_\alpha=\frac{1}{2}(f^2)'(\alpha
A)=\frac{1}{2}f_1''(\alpha A)$, we obtain from \fer{a6}, \fer{a8}:
\begin{eqnarray}
[f^\eta_{1,\alpha},L]&=&-if_\alpha C_1f_\alpha -\alpha f_1''(\alpha
A)C_2-i{\cal R}_{\eta,1}C_1-\frac{\alpha}{2}{\cal R}_{\eta,2} C_2 \nonumber\\
&&+\alpha^2\int_{\r\backslash (-\eta,\eta)} ds\what{f_1}(s) e^{is\alpha
  A}\int_0^sds_1\int_0^{s_1}ds_2\int_0^{s_2}ds_3 \ e^{-is_3\alpha
  A}C_3e^{is_3\alpha A}\nonumber\\
&&-i\alpha^2f_\alpha
\int_\r ds\what{f}(s) e^{is\alpha A}
 \int_0^sds_1\int_0^{s_1}ds_2 e^{-is_2\alpha A} C_3 e^{is_2\alpha
    A}.
\label{a9}
\end{eqnarray}
Consequently, taking into account estimate \fer{40}, we obtain that 
\begin{eqnarray}
\av{i[f^\eta_{1,\alpha},L]}_{g_\nu\psi_n}&=&\av{C_1}_{\psi_{\alpha,\nu,n}}
-\RE \ i\alpha \av{ f''(\alpha A)C_2}_{g_\nu\psi_n} +\RE\av{{\cal
    R}_{\eta,1}C_1}_{g_\nu\psi_n}\nonumber\\
&&-\RE \ i\frac{\alpha}{2}\av{{\cal R}_{\eta,2}C_2}_{g_\nu\psi_n}+
\O{\frac{\alpha^2}{\sqrt{\nu}}}, 
\label{a10}
\end{eqnarray}
as we show next. 
We have taken the real part on the right side, since the left side is a real
number. To estimate the remainder term, we use condition \fer{40} to
obtain 
\begin{equation*}
\left\| e^{-is_3\alpha A}C_3 e^{is_3\alpha A}g_\nu\psi_n\right\|\leq
k\frac{1}{\sqrt{\nu}}e^{\alpha k'|s_3|},
\end{equation*}
uniformly in $n$, so the middle line in \fer{a9} is estimated from above by
\begin{equation*}
k\frac{\alpha^2}{\sqrt{\nu}}\int_\r ds\ |\what{f_1}(s)|\ |s|^3 e^{\alpha
  k'|s|}\leq k\frac{\alpha^2}{\sqrt{\nu}} e^{\alpha k' K}\int_\r ds\ 
|\what{f_1}(s)|\ |s|^3,
\end{equation*}
where $K<\infty$ is such that $\supp\what{f_1}\subset [-K,K]$. The
exponential is 
bounded uniformly in $0\leq \alpha<1$, hence the r.h.s. is $\leq
k\frac{\alpha^2}{\sqrt{\nu}}$. The last line in \fer{a9} is analyzed in the
same way and \fer{a10} follows. \\
\indent
Finally, we observe that 
\begin{eqnarray*}
\lefteqn{-\RE\av{i\alpha f''(\alpha
 A)C_2}_{g_\nu\psi_n}=-\frac{\alpha}{2}\av{i[f''(\alpha
 A),C_2]}_{g_\nu\psi_n}}\\ 
&=&-\frac{\alpha^2}{2}\av{\int_\r ds\ \what{f''}(s)e^{is\alpha A}\int_0^s ds_1 \ 
 e^{-is_1\alpha A} C_3 e^{is_1\alpha
 A}}_{g_\nu\psi_n}=\O{\frac{\alpha^2}{\sqrt{\nu}}}, 
\end{eqnarray*}
where we use \fer{40} again, as above. A similar estimate yields
\begin{equation*}
\RE \ i\frac{\alpha}{2}\av{{\cal
    R}_{\eta,2}C_2}_{g_\nu\psi_n}=-i\frac{\alpha}{4}\av{[{\cal R}_{\eta,2},
    C_2]}_{g_\nu\psi_n}= \O{\frac{\alpha^2\eta}{\sqrt{\nu}}},
\end{equation*}
and using the bound \fer{40'}, we have that 
\begin{equation*}
\av{{\cal R}_{\eta,1}C_1}_{g_\nu\psi_n}=\O{\frac{\eta}{\nu^p}}.
\end{equation*}
Combining this with \fer{a10} and \fer{a0} shows that 
\begin{equation}
\left|\av{C_1}_{\psi_{\alpha,\nu,n}}\right|\leq
k\left(\frac{\sqrt{\nu}+o(n)}{\alpha} + \frac{\alpha^2}{\sqrt{\nu}}+\frac{\eta}{\nu^p}\right).
\label{dagger}
\end{equation}
 Notice that 
\begin{equation*}
C_1\psi_{\alpha,\nu,n}=\int ds\ \what{f}(s) C_1 e^{is\alpha
  A}g_\nu\psi_n\rightarrow C_1\psi_{\alpha,\nu},
\end{equation*}
as $n\rightarrow \infty$, where $\psi_{\alpha,\nu}:=f_\alpha g_\nu\psi$. This
follows from the boundedness condition \fer{40'} and from
$\psi_n\rightarrow\psi$, $n\rightarrow\infty$, see \fer{*}. Consequently we
obtain by taking the limit $n\rightarrow\infty$ in \fer{dagger}
\begin{equation*}
\left|\av{C_1}_{\psi_{\alpha,\nu}}\right|\leq
k\left(\frac{\sqrt{\nu}}{\alpha}+ \frac{\alpha^2}{\sqrt{\nu}}+\frac{\eta}{\nu^p}\right).
\end{equation*}
Choose for instance $\nu=\alpha^3$, $\eta=\alpha^{3p+\delta}$, for any
  $\delta>0$, then 
\begin{equation*}
\lim_{\alpha\rightarrow 0} \av{C_1}_{\psi_{\alpha,\alpha^3}} =0.
\end{equation*}
This concludes the proof of the theorem.\hfill $\blacksquare$
\ \\

{\it Proof of Theorem \ref{virialthm'}.\ }
We adopt the definitions and notation introduced in the proof of Theorem
\ref{virialthm}. It suffices to prove
\begin{equation*}
\lim_{\alpha\rightarrow 0}\scalprod{\psi_\alpha}{i[L,A_0]\psi_\alpha}=0,
\end{equation*}
where we set $\psi_\alpha=\psi_{\alpha,\nu}|_{\nu=\alpha^3}$; see in the proof
of Theorem \ref{virialthm}. The scalar product can be estimated by
\begin{eqnarray*} 
\left|\scalprod{\psi_\alpha}{i[L,A_0]\psi_\alpha}\right|&\leq& 2\left|
  \scalprod{(L-e)\psi_\alpha}{A_0\psi_\alpha}\right|\\
& \leq& 2\|P(N\leq
  n_0)(L-e)\psi_\alpha\|\ \|A_0\psi_\alpha\|.
\end{eqnarray*}
We have 
\begin{eqnarray}
P(N\leq n_0)(L-e)\psi_{\alpha,\nu}&=&\lim_{n\rightarrow\infty} P(N\leq
n_0)[L,f_\alpha] g_\nu\psi_n\label{a11}\\
&&+\lim_{n\rightarrow\infty} P(N\leq n_0)f_\alpha[L,g_\nu]\psi_n.\label{a12}
\end{eqnarray}
Using condition \fer{40'}, we easily find (expanding the commutator
$[L,f_\alpha]$) that $\|P(N\leq n_0)[L,f_\alpha] g_\nu\psi_n\|\leq
k_{n_0}\alpha$. Similarly, using 
\fer{41}, we find that $\| P(N\leq n_0)f_\alpha[L,g_\nu]\psi_n  \|\leq
k\sqrt{\nu}$. It follows that 
$
\|P(N\leq n_0) (L-e)\psi_{\alpha}\|\leq k_{n_0}\alpha.
$
\hfill $\blacksquare$

\ \\

{\it Proof of Theorem \ref{roethm}.\ } The inequality $C\geq\pp-B$, the
continuity of $B$, and \fer{wtl} imply that for any $\epsilon>0$, there is an
$\alpha_0(\epsilon)$, s.t. if $\alpha<\alpha_0(\epsilon)$ then 
\begin{equation}
\scalprod{\psi_\alpha}{\pp\psi_\alpha}\leq\scalprod{\psi}{B\psi}+\epsilon.
\label{star}
\end{equation}
Let $\mu_\phi$ be the spectral measure of $\pp$ corresponding to some
$\phi\in\h$. Then
\begin{equation*}
\scalprod{\psi_\alpha}{\pp\psi_\alpha}=\int_{\r_+}
pd\mu_{\psi_\alpha}(p)=\lim_{R\rightarrow\infty} \int_{\r_+}p\chi(p\leq
R)d\mu_{\psi_\alpha}(p), 
\end{equation*}
where $\chi(p\leq R)$ is the indicator of $[0,R]$. We obtain from \fer{star}
\begin{equation*}
  \lim_{R\rightarrow\infty}\scalprod{\psi_\alpha}{\chi(\pp\leq
  R)\pp\psi_\alpha}=\lim_{R\rightarrow\infty}\left\|\chi(\pp\leq R)\pp^{1/2}
  \psi_\alpha\right\|^2\leq\scalprod{\psi}{B\psi} +\epsilon \equiv k.
\end{equation*}
We have 
$
\left\|\chi(\pp\leq R)\pp^{1/2} \psi\right\|\leq R^{1/2}\|\psi-\psi_\alpha\|
+\sqrt{k}$, 
and taking $\alpha\rightarrow 0$ gives $\|\chi(\pp\leq
R)\pp^{1/2}\psi\|\leq \sqrt{k}$, uniformly in $R$, so 
$
\lim_{R\rightarrow\infty} \int_0^R pd\mu_{\psi}(p)
$
exists and is finite, by the monotone convergence theorem. Since 
$
\dom(\pp^{1/2})=\left\{\psi\left| \int_0^\infty
    pd\mu_\psi(p)<\infty\right.\right\},
$
we have that $\psi\in\dom(\pp^{1/2})$, and 
$\|\pp^{1/2}\psi\|\leq\scalprod{\psi}{B\psi}$.  \hfill $\blacksquare$

\section{Flows and induced unitary groups}

\label{flowsection}
Let $R\subseteq \r^n$ be a Borel set of $\r^n$ (with non-empty interior),  let
$X$ be a  vector 
field on $\r^n$, and consider the initial value problem for $x\in R$:
\begin{eqnarray}
\frac{d}{dt}x_t&=&X(x_t),\nonumber\\
x_t|_{t=0}&=& x. \label{a13}
\end{eqnarray}
We assume that $X$ has the property that, for any initial condition $x\in R$,
there is a unique, 
global (for all $t\in\r$) solution $x_t\in R$ to \fer{a13}. Let $\Phi_t$
denote the corresponding flow  and
assume $\Phi_t$ is a diffeomorphism of $R$ into $R$, for all $t\in\r$. The  
following properties of the flow will be needed:
$\Phi_{s+t}=\Phi_s\circ\Phi_t$,  
$\Phi^{-1}_t=\Phi_{-t}$, $\Phi_0=\bbbone$. The Jacobian determinant of
$\Phi_t(x)$ is given by 
\begin{equation}
J_t(x)=\left|\det\Phi_t'(x)\right|,
\label{a14}
\end{equation}
where $\Phi'_t(x)$ is the matrix $\left(\frac{\partial(\Phi_t)_i}{\partial
    x_j}(x)\right)$. \\
\indent
Let $\mu:R\rightarrow \r_+$ be a continuous function which is $C^1$ on the
    interior of $R$ and which is strictly positive except possibly on a set of
    measure zero. We
 write $d\mu$ for the absolutely continuous measure $\mu(x)dx$, where
    $dx$ denotes Lebesgue measure on $\r^n$. Given a Hilbert space $\hh$,
    consider 
    $L^2(R,d\mu;\hh)$, the space of square integrable functions
    $\psi:R\rightarrow 
    \hh$, equipped with the scalar product
\begin{equation*} 
\scalprod{\psi}{\phi}=\int_R \scalprod{\psi(x)}{\phi(x)}_{\hh}d\mu(x).
\end{equation*}
On the Hilbert space $L^2(R,d\mu;\hh)$, the flow $\Phi_t$ induces a
strongly continuous unitary group, $U_t$, defined by
\begin{equation}
(U_t\psi)(x)=\sqrt{J_t(x)\frac{\mu(\Phi_t(x))}{\mu(x)}}\psi(\Phi_t(x)),
\label{a15}
\end{equation}
for $\psi\in L^2(R,d\mu;\hh)$. 
To check that $U_t$ preserves the norm, we make the change of variables
$y=\Phi_t(x)$ to arrive at
\begin{eqnarray*}
  \int_R \left|(U_t\psi)(x)\right|^2d\mu(x)&=&\int_R J_t(x)|\psi(\Phi_t(x))|^2
  \mu(\Phi_t(x))dx\\
&=&\int_R J_t(\Phi_t^{-1}(y)) |\det(\Phi_t^{-1})'(y)|\ |\psi(y)|^2 \mu(y) dy.
\end{eqnarray*}
We observe that 
$
J_t(\Phi_t^{-1}(y)) \ |\det(\Phi_t^{-1})'(y)|= |\det \bbbone| =1,
$
hence $\|U_t\psi\|=\|\psi\|$. Next, using that
$\Phi_{t+s}=\Phi_t\circ\Phi_s$, one easily shows that
$J_{t+s}(x)=J_t(\Phi_s(x)) J_s(x)$, and that 
\begin{equation*}
\frac{\mu(\Phi_{t+s}(x))}{\mu(x)}=
\frac{\mu(\Phi_t(\Phi_s(x)))}{\mu(\Phi_s(x))}\ \frac{\mu(\Phi_s(x))}{\mu(x)},
\end{equation*}
hence $t\mapsto U_t$ is a unitary group.\\
\indent
In order to see that the unitary group is strongly 
continuous and to calculate its generator, we impose some
additional assumptions on $\mu$ and $X$.  
\begin{itemize}
\item[1.] $X$ is $C^\infty$ and bounded,
\item[2.] for any compact set $M\subset R$, there is a $k<\infty$ s.t. $\partial_t|_{t=0} J_t(x)\leq
  k$, uniformly in $x\in M$,
\item[3.] for any compact set $M\subset R$, there is a $k<\infty$
  s.t. $\left\|\frac{X'(x)\nabla\mu(x)}{\mu(x)}\right\|\leq k$,
  uniformly in $x\in M$,
\item[4.] $t\mapsto \left\{J_t(x)\mu(\Phi_t(x))\right\}^{1/2}$ is $C^1$ in a
  neighbourhood $(-t_0,t_0)$ of zero, and for any compact set $M\subset R$,
  there 
  is a $k<\infty$ s.t. we have the estimate
  $\left|\{J_t(x)\mu(\Phi_t(x))\}^{1/2}\right|<f(x)$, for $|t|<t_0$, where
  $f\in L^2_{\rm loc}(R,dx)$. 
\end{itemize}
If $X$ is $C^\infty$ then so is $\Phi_t(x)$ (jointly in $(t,x)$),
and using that 
\begin{eqnarray}
\Phi_t(x)&=&x+\int_0^t X(\Phi_s(x))ds,\nonumber\\
\Phi_t'(x)&=& \bbbone+\int_0^t X'(\Phi_s(x))\Phi_s'(x)ds,\label{mmm}
\end{eqnarray} 
it follows immediately that
\begin{equation}
\|\Phi_t(x)\|\leq \|x\|+|t|\,\|X\|_\infty,
\label{a24}
\end{equation}
where the subscript $_\infty$ denotes the supremum norm over $x\in R$. In
order to obtain an estimate on $\|\Phi_t'(x)\|$ (the operator norm on ${\cal
  B}(\r^n)$, i.e. the matrix norm, for $x$ fixed), we recall Gronwall's
Lemma. If $\mu:\r\rightarrow \r_+$ is continuous, and $\nu:\r\rightarrow \r_+$
is locally integrable, then the inequality
\begin{equation*}
\mu(t)\leq c +\int_{t_0}^t\nu(s)\mu(s)ds,
\end{equation*}
where $c\geq 0$, and $t\geq t_0$, implies that 
\begin{equation}
\mu(t)\leq c e^{\int_{t_0}^t\nu(s)ds}.
\label{gr}
\end{equation}
Equation \fer{mmm} implies
\begin{equation*}
\|\Phi_t'(x)\|\leq 1+\|X'\|_\infty\int_0^t \|\Phi_s'(x)\| ds,
\end{equation*}
so Gronwall's Lemma yields the estimate
\begin{equation*}
\|\Phi_t'(x)\|\leq \exp(\|X'\|_\infty t),\ \ \ \forall t\geq 0.
\end{equation*}
A similar bound holds for $t<0$, and hence
\begin{equation}
\|\Phi_t'(x)\|\leq \exp \left(\|X'\|_\infty\, |t|\right),\ \ \ \ \ t\in\r,
\label{a25}
\end{equation}
from which it follows that 
\begin{equation}
J_t(x)\leq \exp \left(n\|X'\|_\infty\,|t|\right).
\label{a26}
\end{equation}

For $\psi\in C_0^\infty$, 
\begin{eqnarray}
-\frac{1}{i}\partial_t|_{t=0}(U_t\psi)(x)&=&-\frac{1}{i}\left(\frac{1}{2}\partial_t|_{t=0}J_t(x)+\frac{1}{2}\frac{\nabla\mu(x)\cdot
    X(x)}{\mu(x)} +X(x)\cdot\nabla\right)\psi(x)\nonumber\\
&=&(A\psi)(x),
\label{a21}
\end{eqnarray}
which defines an operator $A$ on $C^\infty_0$. Notice that due to conditions
1, 2 and 3, $A$ maps $C^\infty_0$ into $L^2(R,d\mu;\hh)$. 

\begin{proposition}
\label{generatorlemma}
\stepcounter{theorem}
Assume conditions 1-4 hold. Then for any $\psi\in C^\infty_0$, in the
strong sense on $L^2(R,d\mu;\hh)$,
\begin{equation}
-\frac{1}{i}\frac{U_t-\bbbone}{t}\psi\longrightarrow A\psi,\ \ \
t\rightarrow 0.
\label{a22}
\end{equation}
Consequently, $C_0^\infty$ is in the domain of definition of the selfadjoint
generator of the unitary group $U_t$, and on $C_0^\infty$,
this generator can be identified with the operator $A$ of equation \fer{a21}. 
\end{proposition}

{\it Proof.\ } Invoking the dominated convergence theorem, it is enough to
verify that 
\begin{equation}
\left\|-\frac{1}{i}\frac{1}{t}(U_t-\bbbone)\psi(x)-(A\psi)(x)\right\|^2_\hh
\label{a23}
\end{equation}
is bounded above by a $d\mu$-integrable function which is independent of $t$,
for small $t$. We write
\begin{eqnarray}
\fer{a23}&\leq&\frac{1}{\mu(x)}\left|\frac{1}{t}\left(\sqrt{J_t(x)\mu(\Phi_t(x))}
    -\sqrt{\mu(x)}\right)\right|^2\|\psi(\Phi_t(x)\|^2_\hh\label{a27}\\
&&+\frac{1}{t^2}\|\psi(\Phi_t(x))-\psi(x)\|^2_\hh\label{a28}\\
&&+\|(A\psi)(x)\|^2_\hh.\label{a29}
\end{eqnarray}
Clearly, \fer{a29} is integrable, and, using the continuity properties of
$\psi$ and $\Phi$ and the bound \fer{a25}, one sees that 
\fer{a28} is bounded above by a $t$-independent function that is
$d\mu$-integrable (use the mean value theorem). Next, if $\psi$ has support in a ball of radius $\rho$ in
$R\subset\r^n$, then $\psi\circ\Phi_t$ has support in the ball of radius
$\rho+|t|\, \|X\|_\infty\leq \rho+\|X\|_\infty$, for $|t|\leq 1$. This follows
from 
\fer{a24}. Let $\chi(x)$ denote the indicator function on the ball of radius
$\rho+\|X\|_\infty$, then we have for $|t|<t_0$ with $t_0$ as in condition 4, 
\begin{equation*}
\fer{a27}\leq
k\chi(x)\frac{1}{\mu(x)}\left|\frac{1}{t}\left(\sqrt{J_t(x)\mu(\Phi_t(x))}-\sqrt{\mu(x)}\right)\right|^2\leq
  k\chi(x)\frac{1}{\mu(x)}|f(x)|^2,
\end{equation*}
where we have used the mean value theorem and condition 4. The latter function
is $d\mu$-integrable.\hfill $\blacksquare$
\ \\

{\it Proof of Proposition \ref{xiprop}.\ } Since $\xi$ is globally Lipshitz (with Lipshitz constant
$\|\xi'\|_\infty$),  we have existence and uniqueness of global solutions
to the initial value problem \fer{a13}. Due to uniqueness and the fact that
$\r\ni t\mapsto e_t=0$ is a solution (since $\xi(0)=0$), we see that
$\Phi_t(e)\in (0,\infty)$, for all $t\in\r, e\in (0,\infty)$, so $\Phi_t$ is a
diffeomorphism in $\r_+$. It is not difficult to verify that conditions
 1-4 above are satisfied. Consequently, it
follows from Proposition \ref{generatorlemma} that $C_0^\infty\subset\dom(A)$,
and 
that $A$ is given by \fer{a33} on $C_0^\infty$. Since $\xi$ is infinitely many
times differentiable, $A$ leaves $C_0^\infty$ invariant. Hence
$C_0^\infty$ is a core for $A$.\hfill $\blacksquare$

\section{Proofs of some propositions}
\label{propproofsection}

{\it Proof of Proposition \ref{fgrprop}.} 
Since $\Pi I\Pi=0$ and $\Pi I\repsilon^2 (\pbar_0\otimes\pbar_0) I\Pi=0$, we
have
\begin{eqnarray}
\Pi I\repsilonbar^2 I\Pi=\Pi I\repsilon^2 I\Pi&=&\Pi
I\repsilon^2(\pbar_0\otimes p_0+p_0\otimes\pbar_0) I\Pi\nonumber\\
&& + \Pi I\repsilon^2 (p_0\otimes p_0)I\Pi.
\label{1.1}
\end{eqnarray}
It is not difficult to see that $\epsilon \Pi I\repsilon^2 (p_0\otimes
p_0) I\Pi\rightarrow 0$, as $\epsilon\rightarrow 0$, so the last term in
\fer{1.1} does not contribute effectively to a lower bound in the limit
$\epsilon\rightarrow 0$. 
Let $J$ be the modular conjugation
operator introduced in \fer{J}. Using the relations $J^2=J$,
$Jp_0\otimes\pbar_0= \pbar_0\otimes p_0 J$, 
$J\repsilon^2=\repsilon^2 J$, $JI=-IJ$, and the invariance of 
$\varphi_0\otimes\varphi_0\otimes\Omega$ under $J$, one verifies easily that 
\begin{eqnarray*}
\lefteqn{
\Pi I \repsilon^2 (p_0\otimes \pbar_0) I\Pi=
\Pi I \repsilon^2 (\pbar_0\otimes p_0) I\Pi}\\
&=&\sum_{\alpha,\alpha'}\Pi\big( G_\alpha\otimes\bbbone_p\otimes
  a(\tau_\beta(g_\alpha))\big) \frac{\pbar_0\otimes
  p_0}{(H_p\otimes\bbbone_p -E+L_f)^2 +\epsilon^2} \\
&&\ \ \times
\big( G_{\alpha'}\otimes\bbbone_p\otimes
  a^*(\tau_\beta(g_{\alpha'}))\big)\Pi,
\end{eqnarray*}
where $L_f=\d\Gamma(u)$ and where $\tau_\beta$ has been defined in \fer{9}. 
We pull the annihilation operator through the resolvent, using the pull
through-formula (for $f\in L^2(\r\times S^2)$)
\begin{equation*}
a(f) L_f=\int_{\r\times S^2} \overline{f}(u,\Sigma)  (L_f+u)a(u,\Sigma),
\end{equation*}
and then contract it with the creation operator. This gives the bound 
\begin{eqnarray*}
\Pi I\repsilonbar^2 I\Pi&\geq& \int_{-\infty}^E
du\int_{S^2}d\Sigma \ \frac{u^2}{e^{-\beta u}-1}\\
&&\times \left(p_0 F(-u,\Sigma) \frac{\pbar_0}{(H_p-E+u)^2+\epsilon^2}
  F(-u,\Sigma)^*p_0\right)\otimes p_0\otimes P_\Omega,
\end{eqnarray*}
where we restricted the domain of integration over $u$ to $(-\infty,E)\subset
\r_-$ (as $\epsilon\rightarrow 0$, $\frac{\epsilon}{(H_p-E+u)^2+\epsilon^2}$
tends to the Dirac distribution $\delta(H_p-E+u)$, hence $u=-H_p+E\in
(-\infty, E)$), and where we used \fer{9}. The desired result \fer{fgrc1} now
follows by making the change of variable $u\mapsto -u$ in the integral, and
by remembering the definition of $\gamma$, \fer{FGRC}.
\hfill $\blacksquare$\ \\

{\it Proof of Proposition \ref{alemma}.}\ 
First, we prove a bound on $\Lambda_p e^{itA_p^a}\psi$, for $\psi\in
C_0^\infty$. Let 
$\Phi^a_t$ denote the flow generated by the vector field $\xi_a$. Then, for
each $e\in [0,\infty)$, 
$((\Lambda_p-\bbbone_p) e^{itA_p^a}\psi)(e)=e\psi(\Phi_t^a(e))$, and
\begin{eqnarray}
\left\| (\Lambda_p-\bbbone_p)e^{itA_p^a}\psi\right\|^2&=&\int_{\r_+}
e^2\left\|\psi(\Phi_t^a(e))\right\|^2 de\nonumber\\
&=&\int_{\r_+}\left(\Phi_{-t}^a(y)\right)^2\|\psi(y)\|^2\left(\Phi_{-t}^a\right)'(y)dy,
\label{b1}
\end{eqnarray}
where we make the change of variables $y=\Phi_t^a(e)$. Recall that
$\Phi_t^a(y)=y+\int_0^t\xi(\Phi_s^a(y)/a) ds$, so 
\begin{equation}
|\Phi_t^a(y)|\leq |y|+ |t|\, \|\xi\|_\infty.
\label{b2}
\end{equation}
Next $(\Phi_t^a)'(y)=1+\int_0^t\frac{1}{a}\xi'(\Phi_s^a(y)/a)\,
(\Phi_s^a)'(y)ds$ yields 
\begin{equation}
\left|(\Phi_t^a)'(y)\right|\leq
1+\int_0^t\frac{1}{a}\|\xi'\|_\infty\left|(\Phi_s^a)'(y)\right| ds,
\label{b3}
\end{equation}
and Gronwall's estimate, \fer{gr}, implies that
\begin{equation}
|(\Phi_t^a)'(y)|\leq e^{\|\xi'\|_\infty |t|/a}.
\label{b5}
\end{equation}
Using \fer{b5} and \fer{b2} in \fer{b1} yieds 
\begin{eqnarray*}
\left\| (\Lambda_p-\bbbone_p)e^{itA_p^a}\psi\right\|^2 &\leq& 
e^{\|\xi'\|_\infty 
  |t|/a}\int_{\r_+} \left( y+\|\xi\|_\infty|t|\right)^2 \|\psi(y)\|^2
dy\nonumber\\
&\leq&  2e^{\|\xi'\|_\infty |t|/a}\left(1+\|\xi\|_\infty|t|\right)^2\left(
  \|(\Lambda_p-\bbbone_p)\psi\| +\|\psi\|\right)^2,
\end{eqnarray*}
from which it follows that 
\begin{eqnarray}
\left\| \Lambda_pe^{itA_p^a}\psi\right\| &\leq& 4\sqrt{2}
\left(1+\|\xi\|_\infty|t| \right)e^{\|\xi'\|_\infty |t|/a}
\|\Lambda_p\psi\|\nonumber \\
&\leq& 4\sqrt{2} e^{(\|\xi'\|_\infty+\|\xi\|_\infty)|t|/a}\|\Lambda_p\psi\|.
\label{b6}
\end{eqnarray}
Estimate \fer{b6} holds for all $\psi\in C_0^\infty$, which is a core for
$\Lambda_p$. Next, let $\psi\in\dom(\Lambda_p)$, and let $\{\psi_n\}\subset
C_0^\infty$ be a sequence, s.t. $\psi_n\rightarrow\psi$,
$\Lambda_p\psi_n\rightarrow \Lambda_p\psi$, as $n\rightarrow \infty$. If
$\chi_R$ denotes the cutoff function $\chi(\Lambda_p\leq R)$, for $R>0$,
we have
\begin{eqnarray*}
\left\|\chi_R\Lambda_p e^{itA_p^a}\psi\right\|&\leq& \left\|\chi_R \Lambda_p
  e^{itA_p^a}\psi_n\right\| +R\|\psi-\psi_n\|\\
&\leq& 4\sqrt{2} e^{(\|\xi'\|_\infty+\|\xi\|_\infty)|t|/a}
\|\Lambda_p\psi_n\| +R\|\psi-\psi_n\|.
\end{eqnarray*}
Taking $n\rightarrow\infty$ yields 
\begin{equation*}
\left\|\chi_R\Lambda_p e^{itA_p^a}\psi\right\|\leq
4\sqrt{2} e^{(\|\xi'\|_\infty+\|\xi\|_\infty)|t|/a}
\|\Lambda_p\psi\|,
\end{equation*}
uniformly in the cutoff parameter $R$. This shows that
$e^{itA_p^a}\psi\in\dom(\Lambda_p)$, and \fer{b6} is valid for all
$\psi\in\dom(\Lambda_p)$. \\
\indent
We complete the proof of the proposition by examining
$\Lambda_fe^{itA_f}\psi$. Let 
 $\psi\in\dom_f$. Then one finds the following bound for the
$n$-particle component: 
\begin{eqnarray*}
\left\|\left[(\Lambda_f-\bbbone_f) e^{itA_f}\psi\right]_n\right\|^2&=& \left\|\sum_{j=1}^n
      (u_j^2+1)\psi_n(u_1-t,\ldots,u_n-t)\right\|^2\nonumber\\
&=&\left\|\sum_{j=1}^n\left(
      (u_j+t)^2+1\right)\psi_n(u_1,\ldots,u_n)\right\|^2\nonumber\\
&\leq&\left(
      2(1+t^2)\right)^2\left\|\sum_{j=1}^n(u_j^2+1)\psi_n(u_1,\ldots,u_n)\right\|^2. 
\end{eqnarray*}
It follows that $\|(\Lambda_f-\bbbone_f)e^{itA_f}\psi\|\leq
2(1+t^2)\|\Lambda_f\psi\|$, 
for all $\psi\in\dom_f$, so that 
\begin{equation*}
\|\Lambda_f e^{itA_f}\psi\|\leq 2(1+t^2)\|\Lambda_f\psi\|+\|\psi\|\leq
3e^t\|\Lambda_f\psi\|,
\end{equation*}
for all $\psi\in\dom_f$. A similar argument as above shows that this estimate
extends to all $\psi\in\dom(\Lambda_f)$. 
\hfill $\blacksquare$
\ \\

{\it Proof of Proposition \ref{boundedcomms}.}\ 
We denote the fiber of $A^a_p$ by $A_p^a(e)$, i.e.
\begin{equation}
A_p^a(e)=i\left(\frac{1}{2}\frac{1}{a}\xi'(e/a) +\xi(e/a)\partial_e\right),
\label{fiber} 
\end{equation}
see also  \fer{a36}. For $\psi\in C_0^\infty$,  we have 
\begin{eqnarray*}
(A_p^a G_\alpha\psi)(e)&=&A_p^a(e)(G_\alpha\psi)(e)\\
&=&A_p^a(e)G_\alpha(e,E)\psi(E) +A_p^a(e)\int_{\r_+} G_\alpha(e,e')\psi(e')
de'.
\end{eqnarray*}
Due to the regularity property \fer{regul2}, we 
can take the operator $A_p^a(e)$ inside the integral (Dominated Convergence
Theorem), and obtain the estimate
\begin{eqnarray}
\|A_p^aG_\alpha\psi\|^2
&\leq& |\psi(E)|^2 \int_{\r_+} \|A_p^a(e)G_\alpha(e,E)\|^2_\hh
de\label{a51}\\ 
&&+\int_{\r_+}\left[\int_{\r_+} \|A_p^a(e)G_\alpha(e,e')\psi(e')\|_\hh de'\right]^2de.
\label{a52}
\end{eqnarray}
Using \fer{fiber} and the bound
$|a^{-1}\xi'(e/a)|\leq e^{-1}\sup_{e\geq 0}e\xi'(e)\leq ke^{-1}$, it is easily
seen that the 
integrand of \fer{a51} is bounded above by 
\begin{equation*}
k\left(\|e^{-1}G_\alpha(e,E)\|^2_\hh +
\|\partial_1 G_\alpha(e,E)\|^2_\hh\right),  
\end{equation*}
which is integrable, due to \fer{regul1}. We estimate the integrand in
\fer{a52} by
\begin{eqnarray*}
\lefteqn{
\|A^a_p(e)G_\alpha(e,e')\psi(e')\|_\hh}\\
&&\leq k\left(
  \|e^{-1}G_\alpha(e,e')\|_{{\cal B}(\hh)}+ \|\partial_1
  G_\alpha(e,e')\|_{{\cal B}(\hh)}\right) \|\psi(e')\|_\hh,
\end{eqnarray*}
and using H\"older's inequality, we arrive at
\begin{eqnarray*}
\lefteqn{
\fer{a52}\leq k\int_{\r_+}\|\psi(e)\|^2de}\\
&&\times  \int_{\r_+}\int_{\r_+}
\left\{\|e^{-1}G_\alpha(e,e')\|^2_{{\cal B}(\hh)} + \|\partial_1
  G_\alpha(e,e')\|^2_{{\cal B}(\hh)}\right\} de de'.
\end{eqnarray*}
By condition \fer{regul2}, the double integral is finite. We conclude that 
\begin{equation}
\|A_p^a G_\alpha\psi\|\leq k\|\psi\|.
\label{a53}
\end{equation}
One also finds that $\|G_\alpha A_p^a\psi\|\leq k\|\psi\|$, e.g. by noticing
that $\|G_\alpha A_p^a\psi\|=\sup_{0\neq\phi\in
  C_0^\infty}\|\phi\|^{-1}|\scalprod{\phi}{G_\alpha
  A_p^a\psi}|=\sup_{0\neq\phi\in
  C_0^\infty}\|\phi\|^{-1}|\scalprod{A_p^aG_\alpha\phi}{\psi}|$ and using
\fer{a53}. Consequently, we have shown \fer{a48} for $n=1$. \\
\indent
The proof for $n=2,3$ follows the above lines. For instance, in order to show
boundedness of the third multi-commutator, a typical term to estimate is
$\|A_p^aA_p^aG_\alpha A_p^a\psi\|$, for $\psi\in C_0^\infty$. We shall sketch
the proof that this term is bounded, all other ones being treated
similarly. We have   
\begin{equation}
\|A_p^aA_p^aG_\alpha A_p^a\psi\|=\sup_{0\neq\phi\in
  C_0^\infty}\|\phi\|^{-1}\left|\scalprod{\phi}{A_p^aA_p^aG_\alpha A_p^a\psi}\right|,
\label{a54}
\end{equation}
and the scalar product equals
\begin{equation}
\int_{\r_+}\int_{\r_+}\scalprod{\phi(e)}{A_p^a(e)^2G_\alpha(e,e')A_p^a(e')\psi(e')}_\hh
de\,de'.
\label{a55'}
\end{equation}
Recalling \fer{fiber}, one can calculate the operator
$A_p^2(e)^2G_\alpha(e,e')$. It can be written as a sum of terms, involving
multiplications by functions with argument $e$, and derivatives $\partial_1G_\alpha(e,e')$,
$\partial_1^2G_\alpha(e,e')$. Using the formulas for the adjoints of derivatives of
$\partial_1^{1,2}G_\alpha(e,e')$, see \fer{a56}, we obtain
$[A_p^a(e)^2G_\alpha(e,e')]^*$, and \fer{a55'} becomes
\begin{equation}
\int_{\r_+}\int_{\r_+}\scalprod{A_p^a(e')[A_p^a(e)^2G_\alpha(e,e')]^*\phi(e)}{\psi(e')}_\hh
de\,de',
\label{a55}
\end{equation}
due to selfadjointness of $A_p^a(e')$ on $\hh$, and the fact that for
all $e\in\r_+$,
\begin{equation*}
[A_p^a(e)^2G_\alpha(e,e')]^*\phi(e)\in\dom (A_p^a(e')),
\end{equation*}
which follows from condition \fer{regul2}. Moreover, the same condition allows
us to estimate
\begin{eqnarray*}
\lefteqn{
|\fer{a55}|}\\
&\leq&
\int_{\r_+}\int_{\r_+}\left\||A^a_p(e')[A_p^a(e)^2G_\alpha(e,e')]^*\right\|_{{\cal
    B}(\hh)} \|\phi(e)\|_\hh\|\psi(e')\|_\hh
de\,de'\\
&\leq& \|\phi\|\ \|\psi\| \left[\int_{\r_+}\int_{\r_+}\left\||A^a_p(e')[A_p^a(e)^2G_\alpha(e,e')]^*\right\|_{{\cal
    B}(\hh)}^2de\,de'\right]^{1/2}\\
&\leq&k\|\phi\|\ \|\psi\|, 
\end{eqnarray*}
where we have used H\"older's inequality. This shows that
$\fer{a54}\leq k\|\psi\|$. \hfill $\blacksquare$
\ \\

{\it Proof of Proposition \ref{vtapply}.}\ 
We have mentioned before \fer{a} that $A_0$ satisfies the conditions of
Theorem \ref{virialthm'}, so it suffices to verify the conditions of Theorem
\ref{virialthm}. \\
\indent
We need to check that $(X,\Lambda,\dom)$ is a GJN triple, for $X=L,N,D,C^a_n$,
$n=1,2,3$, and that \fer{alambda}, \fer{41}, \fer{40},\fer{40'} are
satisfied. Proposition \ref{alemma} shows that \fer{alambda} holds. The
operator 
$D$, given in \fer{operatorD}, is clearly $N^{1/2}$-bounded in the sense of
Kato on $\dom$, since $G_\alpha$ are bounded operators, and
$\tilde{g}_\alpha$, $e^{-\beta u/2}\tilde{g}_\alpha$ are
square-integrable. Hence \fer{41} holds. Recalling Remark 1) after Theorem
\ref{virialthm}, and 
noticing that $N$ commutes with $e^{itA^a}$, in the strong sense on $\dom$,
and that $C_3^a\leq kN^{1/2}$, in the sense of Kato on $\dom$ (see \fer{a45}),
we see that \fer{40} is verified. Similarly, $C_1^a\leq kN$ in the sense of
Kato on $\dom$, see \fer{a43}, so \fer{40'} holds.\\
\indent
It remains to show that the above mentioned triples satisfy the GJN
properties. We first look at $(L,\Lambda,\dom)$. Clearly, $\|L\psi\|\leq
k\|\Lambda\psi\|$, for $\psi\in\dom$. Moreover, $L_0$ commutes with $\Lambda$
in the strong sense on $\dom$, so we need only consider the interaction term
in the verification of \fer{nc2}. Due to condition \fer{newcond}, we have for
all $\psi\in C_0^\infty$: $\|\Lambda_pG_\alpha\psi\|\leq k\|\psi\|$,
$\|G_\alpha\Lambda_p\psi\|\leq k\|\psi\|$. Consequently, for $\psi\in\dom$:
\begin{eqnarray}
\lefteqn{
\left|\scalprod{G_\alpha\otimes\bbbone_p \otimes\varphi(\tilde{g}_\alpha)\psi}{\Lambda\psi}
 -\scalprod{\Lambda
   \psi}{G_\alpha\otimes\bbbone_p \otimes\varphi(\tilde{g}_\alpha)\psi}\right|  
}\nonumber\\
&\leq& k\|\psi\|\ \|\varphi(\tilde{g}_\alpha)\psi\|\nonumber\\ 
&&
+\left|\scalprod{G_\alpha\otimes\bbbone_p \otimes\varphi(\tilde{g}_\alpha)\psi}{\Lambda_f\psi} 
 -\scalprod{\Lambda_f
   \psi}{G_\alpha\otimes\bbbone_p \otimes\varphi(\tilde{g}_\alpha)\psi}\right|  
\label{towone}\\
&\leq& k\|\psi\|\ \|\varphi(\tilde{g}_\alpha)\psi\| +k\|\psi\|\
\|\varphi((u^2+1)\tilde{g}_\alpha)\psi\| \nonumber\\
&\leq& k\|\psi\|\ \|\Lambda^{1/2}\psi\| \nonumber\\
&\leq& k\left(\|\psi\|^2+\|\Lambda^{1/2}\psi\|^2\right) \nonumber\\
&\leq& k\scalprod{\psi}{(\Lambda+\bbbone)\psi}\nonumber\\
&\leq& 2k\scalprod{\psi}{\Lambda\psi}.\nonumber
\end{eqnarray}
We used in the  third step that
$\varphi(\tilde{g}_\alpha)$ and 
$\varphi((u^2+1)\tilde{g}_\alpha)$ are 
relatively $\Lambda_f^{1/2}$ bounded, in the sense of Kato on $\dom$. This
follows since $(u^2+1)\tilde{g}_\alpha\in L^2(\r\times S^2,du\times
d\Sigma)$, due to conditions \fer{IR} and \fer{UV}. The same estimates hold
for $\bbbone_p\otimes \cc_p G_\alpha \cc_p \otimes\varphi(e^{-\beta u/2}\tilde{g}_\alpha)$, hence we have shown
that $(L,\Lambda,\dom)$ is a GJN triple.\\
\indent
It is clear that $N\leq \Lambda$ in the sense of Kato on $\dom$, and since $N$
commutes with $\Lambda$ in the strong sense on $\dom$, we see immediately that
$(N,\Lambda,\dom)$ is a GJN triple.\\
\indent
Next, consider $(D,\Lambda,\dom)$. Since $D$ has the same structure as $I$,
c.f. \fer{I} and \fer{operatorD}, the proof that $(D,\Lambda,\dom)$ is a
GJN triple goes as the one for $(L,\Lambda,\dom)$. \\
\indent
We examine $(C_n^a,\Lambda,\dom)$, $n=1,2,3$, $a>0$. Recall that the $C_n^a$
are given in \fer{a43}-\fer{a45}. Each $C_n^a$ has a term that acts purely on
the particle space. This term is a bounded multiplication operator that
commutes with $\Lambda$, in the strong sense on $\dom$. Therefore, we need
only show that $(N+\lambda I^a_1,\Lambda,\dom)$, $(I_{2,3}^a,\Lambda,\dom)$
are GJN triples. Since we have shown it for $(N,\Lambda,\dom)$, it suffices to
treat $(I_n^a,\Lambda,\dom)$, $n=1,2,3$, $a>0$. We take the general term in
the sum of \fer{a46}:
\begin{equation*}
X:={\rm ad}_{A_p^a}^{(j)}(G_\alpha)\otimes\bbbone_p\otimes{\rm
  ad}_{A_f}^{(n-j)}\left(\varphi(\tilde{g}_\alpha)\right).
\end{equation*}
Since ${\rm ad}_{A_p^a}^{(j)}(G_\alpha)$ is bounded, $j=1,2,3$ (see
Proposition \ref{boundedcomms}), and 
\begin{equation*}
{\rm ad}_{A_f}^{(n-j)}\left(\varphi(\tilde{g}_\alpha)\right)
=\varphi\left((i\partial_u)^{n-j}\tilde{g}_\alpha\right)
\end{equation*}
is relatively $\Lambda_f^{1/2}$-bounded, in the sense of Kato on $\dom$ (this
follows from $\partial_u^k\tilde{g}_\alpha\in L^2(\r\times S^2)$, $k=1,2,3$,
due to \fer{IR}, \fer{UV}), then it is clear that $\|X\psi\|\leq
k\|\Lambda\psi\|$, $\psi\in \dom$. Next, we verify condition \fer{nc2} as
above in \fer{towone}:
\begin{eqnarray*}
\left|\scalprod{X\psi}{\Lambda\psi}-\scalprod{\Lambda\psi}{X\psi}\right|
&\leq& k\|\psi\|\ \left\|\varphi\left(
    (u^2+1)(i\partial)^{n-j}\tilde{g}_\alpha\right)\psi\right\|\\
&\leq& k\|\psi\|\ \|\Lambda^{1/2}\psi\|,
\end{eqnarray*}
since $(u^2+1)(\partial_u)^k\tilde{g}_\alpha\in L^2(\r\times S^2)$, for
$k=1,2,3$, due to \fer{IR} and \fer{UV}.\hfill $\blacksquare$
\ \\

{\bf Acknowledgements.\ } We thank I.M. Sigal for numerous stimulating
discussions on the subject matter of this paper. M.M. is grateful to him for
some ideas leading to an early version of Theorem \ref{virialthm}.

\end{document}